**Principles of crystal growth of intermetallic and oxide compounds from molten solutions**

I. R. Fisher, M. C. Shapiro and J. G. Analytis

*Geballe Laboratory for Advanced Materials and Department of Applied Physics, Stanford University, CA 94305, and the Stanford Institute for Materials and Energy Sciences, SLAC National Accelerator Laboratory, 2575 Sand Hill Road, Menlo Park, California 94025, USA*

**Abstract**

We present a tutorial on the principles of crystal growth of intermetallic and oxide compounds from molten solutions, with an emphasis on the fundamental principles governing the underlying phase equilibria and phase diagrams of multicomponent systems.

**Contents**

# 1. Introduction

Detailed investigation of the intrinsic physical properties of materials often requires the measurement of single crystal samples. This is especially true in the realm of quantum materials, for which complex interactions can lead to subtle forms of emergent magnetic and electronic properties. At a basic level, single crystals enable determination of the intrinsic anisotropy of such materials, providing detailed information about important terms in the effective Hamiltonian describing the low energy properties. More broadly, crystal growth is also a purifying process. Structural and compositional disorder can profoundly affect the ground state of strongly correlated systems, or mask the signatures of subtle electronic phase transitions. Furthermore, subtle electronic states can exist close to the boundary of competing phases, and precise control of the stoichiometry is a prerequisite both for determining the intrinsic properties of a stoichiometric "parent" compound, and also for continuous control of the composition via chemical substitution. These reasons all motivate the development of well-controlled methods for the growth of high quality single crystals. In some cases, materials of interest to the condensed matter community are already the subject of extensive research in the broader fields of solid state chemistry or materials science, and avenues for crystal growth may already have been developed and studied for their own intellectual merit. However, for the majority of cases the synthesis of these materials has not been studied in such rigor, and consequently they are not broadly available in single crystal form. In these cases, the interested physicist must either invest the time and resources to grow the single crystals for themself, or at least cultivate a collaborative relationship with someone else who does. In either case, an appreciation of the physical principles that undergird the process of crystal growth is valuable.

This tutorial-style article is intended to provide a brief introduction to the principles of crystal growth from a molten solution seen through the lens of phase equilibria. The discussion draws on and largely follows that found in standard textbooks on the subject, including those by Porter and Easterling [1] and Gaskell [2], and is complemented by several practical examples drawn from our own experience. It is, however, not intended to be an extensive description of the science and art of crystal growth, for which other venerated resources exist (for example [3-10]). Our aim is to provide an appreciation of the factors that determine equilibrium phase diagrams, and to show how these diagrams provide a roadmap for single crystal growth of thermodynamically stable phases from a molten solution. For many materials such phase diagrams are not available, but even in that situation a rudimentary understanding of the factors governing phase relations is indispensable for exploratory crystal growth. We apply these concepts specifically to the broad classes of intermetallic and oxide materials, which span a wide range of materials of current interest, but the concepts are general and can be applied to other systems. A range of different techniques can be drawn upon in order to grow crystals of a specific material from a molten solution, but for the purpose of this article we restrict our discussion of practical examples to flux growth via spontaneous nucleation, which is both widely applicable and can also be implemented with minimal investment in infrastructure.

## 2. Basics of Phase Equilibria

### 2.1 *Binary mixtures*

At constant temperature and pressure, equilibrium phases minimize the Gibbs free energy (*G*) of a system, *G* = *E* + *PV* – *TS* = *H* – *TS*. When the free energy of two phases cross there is a phase transition, which can be either first order or continuous. Melting transitions involve a latent heat and are therefore always first order (Figure 1). Construction of the phase diagram for single component systems, for example water, is treated extensively in standard physics text books and we shan't belabor this. Compounds are formed from mixtures of individual elements, requiring description of a multi-component system. The same general principle of minimizing the free energy to determine the equilibrium phase holds, but calculation of this quantity must now include compositional variation of the enthalpy (*H*) and entropy (*S*) for each phase. The associated theoretical description is covered in several classic text books on metallurgy and materials science (for example [1,2 and 11]) but is often a subject that is not taught in the context of condensed matter physics. In the following section these concepts are introduced with the aim of providing a foundation for understanding general features of equilibrium phase diagrams of multicomponent systems. The discussion largely follows that of [1], with several figures inspired from that text.

We first treat the case of a binary mixture of elements *A* and *B* for which both elements have the same crystal structure in the solid phase. The total number of atoms of each component are $N_A$ and $N_B$, and the relative fractions are $X_A = N_A / (N_A + N_B)$ and $X_B = N_B / (N_A + N_B)$ such that $X_A + X_B = 1$. Before mixing, the free energy per mole is given by $G_A$ and $G_B$. Neglecting the effects of mixing, the total free energy per mole would simply be given by $G = X_A G_A + X_B G_B$, reflecting the relative amounts of the two components. However, the process of mixing affects the chemical bonding, changing the enthalpy, and also introduces disorder, changing the entropy. Both of these effects contribute to a difference in the free energy due to the mixing, $\Delta G_{mix}$, relative to the sum given above.

An ideal solution is one for which the enthalpy of mixing, $\Delta H_{mix}$, is zero, as is approximately the case for Au-Ag alloys or other mixtures of very similar elements. In this case, the only contribution to the change in free energy upon mixing $\Delta G_{mix}$ is due to the entropy of mixing, $\Delta S_{mix}$. If there is no change in the temperature or volume of the system on mixing the two components *A* and *B*, then $\Delta S_{mix}$ is due solely to the change in configurational entropy. The number of ways that the two components can form an ideal substitutional solid solution is then given by $W = (N_A + N_B)!/(N_A! N_B!)$ (see Figure 2). The resulting expression for $\Delta S_{mix}$ per mole (setting $N_A + N_B = N_{Av}$, Avogadro's number) can be simplified using Stirling's approximation ($\ln N! \approx N \ln N - N$), yielding the following expression after substitution of the relative molar fractions $X_A$ and $X_B$:

$$\Delta S_{mix} = -R(X_A \ln X_A + X_B \ln X_B) \qquad (1)$$

where *R* is the molar gas constant. This quantity is necessarily positive since both $X_A$ and $X_B$ are less than one. Physically, the mixed system is more disordered than the pure cases of either $X_A = 1$ or $X_B = 1$. The associated change in the free energy due to mixing is therefore negative, and is given by $\Delta G_{mix} = -T\Delta S_{mix}$ (Figure 3). This result holds equally for an ideal solution of two liquids as for two solids (consider a snapshot of the liquid at a given moment in time and apply a distorted lattice such that each atom defines a lattice point; the change in configurational entropy due to mixing is the same as for the solid case.)

The term "regular solution" is used for solutions for which there is only moderate deviation from the behavior of an ideal solution. We now account for changes in the enthalpy due to mixing, but assume that the interactions that result in this do not fundamentally change the entropy of mixing. Assuming only short range nearest neighbor interactions, and also assuming that the configurational entropy is not changed by the process of mixing, it is straightforward to derive the associated enthalpy of mixing in terms of the energy associated with bonds between like atoms ($E_{AA}$ and $E_{BB}$) and between dissimilar atoms $E_{AB}$. For the unmixed case, the internal energy due to chemical bonding is simply $E_{unmixed} = zN_A E_{AA}/2 + zN_B E_{BB}/2 = zN_{av}X_A E_{AA}/2 + zN_{av}X_B E_{BB}/2$, where $z$ is the coordination number and the factor of ½ avoids double counting bonds between pairs of atoms (we assume that the system is large enough that the number of dangling bonds at the surface is negligible compared to the number of satisfied bonds in the bulk). For the mixed case, the number of bonds between two atoms of type A is $P_{AA} = zX_A^2 N_{av}/2$ and similarly between atoms of type B is $P_{BB} = zX_B^2 N_{av}/2$, where the factor ½ again avoids double counting, and the number of bonds between dissimilar atoms is $P_{AB} = zX_A X_B N_{av}$ (see Figure 4). The internal energy of the mixed components due to chemical bonding is then $E_{mixed} = P_{AA}E_{AA} + P_{BB}E_{BB} + P_{AB}E_{AB}$. The change in internal energy due to chemical bonding is given by $\Delta E_{mix} = E_{mixed} - E_{unmixed} = zN_{av}X_A X_B[E_{AB}-(E_{AA} + E_{BB})/2]$, where we have used the relation $X_A + X_B = 1$ to obtain this simplified expression. Hence, at constant volume the enthalpy of mixing is quadratic and is given by

$$\Delta H_{mix} = uX_A X_B \qquad (2)$$

where $u$ is a constant, determined by the difference of $E_{AB}$ and the average value of $E_{AA}$ and $E_{BB}$.

Combining the effects of the entropy of mixing and the enthalpy of mixing we arrive at the free energy change of mixing for a regular solution:

$$\Delta G_{mix} = \Delta H_{mix} - T\Delta S_{mix} = uX_A X_B + RT(X_A \ln X_A + X_B \ln X_B) \qquad (3)$$

Recalling that $\Delta S_{mix} > 0$, mixing results in a decrease in the free energy for all temperatures if the reaction is exothermic ($\Delta H_{mix} < 0$). However, for endothermic reactions ($\Delta H_{mix} > 0$) the competing effects of the two terms in Equation 3 can lead to two distinct minima in $\Delta G_{mix}$ at low temperatures, affecting the miscibility of the two components (Figure 5). Real solutions can be significantly more complex, but the above description provides an excellent starting point for a discussion of phase diagrams.

Before working out the phase diagram for a binary mixture, it is useful to see how the chemical potential $\mu$ varies in such a system. The fundamental thermodynamic equation for a system with a single component and variable amount of material $N$ is $dE = TdS - PdV + \mu dN$, from which $dG = VdP - SdT + \mu dN$, and hence $\mu = (\partial G/\partial N)_{P,T}$. Furthermore, since the Gibbs free energy is an extensive quantity (i.e. is proportional to $N$), we can write $\lambda G(T,P,N) = G(T,P,\lambda N)$. Differentiating with respect to $\lambda$, and then setting $\lambda = 1$, gives $G = N\mu$. These relations generalize to a system with multiple components in a straightforward way. Hence, for a binary system the differential of the *molar* Gibbs free energy is $dG = VdP - SdT + \mu_A dX_A + \mu_B dX_B$, where $\mu_A$ and $\mu_B$ are measured in J/mol. From this we obtain $(\partial G/\partial X_B)_{P,T} = \mu_B - \mu_A$ since $X_A = 1 - X_B$. Similarly, the molar free energy is $G = \mu_A X_A + \mu_B X_B$. Combining these equations we obtain the following two useful expressions for the chemical potential of components *A* and *B* as a function of the composition of the mixture $X_B$:

$$\mu_A = G - X_B(\partial G/\partial X_B)_{P,T} \qquad (4)$$

$$\mu_B = G + (1-X_B)(\partial G/\partial X_B)_{P,T} = \mu_A + (\partial G/\partial X_B)_{P,T} \qquad (5).$$

These expressions are completely general (i.e. independent of the actual functional form of $G(X_B)$), and suggest a simple graphical method to determine the chemical potential of each component in the mixture. Specifically, the chemical potential of each component is given by the intercept of the tangent to the free energy with the vertical axis, as illustrated in Figure 6. Such a graphical construction is often referred to as the *method of intercepts*.

We are now in a position to describe the phase diagram of this simple binary mixture of elements. As a start, consider the case for which the two components are completely miscible and have the same crystal structure in the solid phase. The progression of the compositional dependence of the free energy of both the liquid and solid phases upon cooling the system is illustrated in Figure 7. At sufficiently high temperature the liquid phase is stable for all compositions (Figure 7(a)). Upon cooling, the reduction in the entropy progressively reduces the difference in the free energy of the solid and liquid phases (see Figure 1), and at sufficiently low temperatures the solid phase is stable for all compositions (Figure 7(d)). The most interesting cases correspond to temperatures between the melting points of the two elements $T_m(A)$ and $T_m(B)$ (Figure 7(c)). In this range of temperatures the free energy curves for liquid and solid phases cross, and for compositions between the minima of $G_L$ and $G_S$ it is energetically favorable for the system to phase separate into a mixture of liquid and solid. Two key parameters that we wish to calculate are then the composition of the liquid and solid phases in equilibrium at a given temperature, and the relative molar amounts of the two phases.

To determine the composition of the solid and liquid phases we recall that for a closed system in equilibrium the chemical potential of a component (whether *A* or *B*) must be the same in both phases; i.e. $\mu_A^S = \mu_A^L$ and $\mu_B^S = \mu_B^L$. With reference to the method of intercepts illustrated in Figure 6, to satisfy both requirements requires that the free energy of the equilibrium compositions $G_S(x_S)$ and $G_L(x_L)$ have a *common tangent*, as illustrated in Figure 8(a). Following this procedure, calculation of the equilibrium compositions of the liquid ($x_L$) and solid ($x_S$) phases for each temperature between $T_m(A)$ and $T_m(B)$ results in the phase diagram, shown in Figure 8(b).

The relative amount of the two phases in equilibrium can be calculated using the *lever rule*. Specifically, with reference to Figure 8(b), if the composition of the mixture is given by $X_B = x$ then the relative amount of the solid phase is given by $(x_L-x)/(x_L-x_S)$ while the relative amount of the liquid phase is given by $(x-x_S)/(x_L-x_S)$. In other words, if *x* is close to $x_s$, there is more solid than liquid, and vice versa. Proof of these relations is straight forward and follows from the conservation of the amount of each of the components. We start by specifying the molar amount of the solid phase (*P* moles with composition $x_S$) and the molar amount of the liquid phase ($Q$ moles with composition $x_L$) such that the total amount of component A is given by $N_A = N_A(S) + N_A(L) = P(1-x_S)N_{av} + Q(1-x_L)N_{av}$ and likewise $N_B = N_B(S) + N_B(L) = Px_SN_{av} + Qx_LN_{av}$. Hence $(x_L-x)/(x_L-x_S) = (N_B(L)/QN_{av} - N_B/(N_A+N_B))/(N_B(L)/QN_{av} - N_B(S)/PN_{av})$. Noting that $N_A + N_B = (P + Q)N_{av}$, a little manipulation yields $(x_L-x)/(x_L-x_S) = P/(P+Q)$. A similar manipulation yields the result for the relative amount of the liquid phase, $Q/(P+Q)$.

A simple extension of the ideas introduced above is to consider the case in which the end members of the binary phase diagram have different structures, $\alpha$ and $\beta$. In this case, we must consider the free energy of three distinct phases, $\alpha$, $\beta$ and the liquid *L*. Following exactly the same principles described above, whenever the composition lies between two minima of the free energy it is energetically favorable for the system to phase separate, in this case to either $\alpha + L$, $\beta + L$ or $\alpha + \beta$. The construction of the associated phase diagram proceeds via a similar method of intercepts, and is illustrated in Figure 9 for a representative case. The composition with the minimum melting point is referred to as a *eutectic mixture* (the word deriving the Greek *eutektos*, meaning "easy to melt"). The *eutectic point* is an invariant point on a phase diagram: the system has no degrees of freedom, and hence no independent changes in the state of the system can be made (see Section 2.2).

One can also construct a phase diagram for systems in which several solid phases are possible, an example of which is given in Figure 10. In this particular example, there is still a reasonable solubility of each component in each of the solid phases, and the range of compositions over which each phase is stable is quite considerable. Intermediate phases can be either *congruently melting* (melt from a homogeneous solid to a homogeneous liquid) or *incongruently melting* (the solid decomposes on heating to a two-phase mixture of solid and liquid phases each with a different composition to the original solid). The temperature at which such a decomposition occurs is referred to as the *peritectic temperature*. More generally, a *peritectic reaction* is one in which a solid and liquid transform in to a different solid ($\alpha + L \rightarrow \beta$) and vice versa. (In this case, the word derives from the Greek *peri* & *tektikos*, meaning "able to dissolve or melt".)

Looking beyond regular solutions, large negative departures from ideality can occur due to chemical bonding. In these cases, large negative values of $\Delta H_{mix}$ (i.e. strongly exothermic reactions) occur for specific compositions which satisfy bonding requirements. In these cases the large gain in internal energy due to chemical bonding far outweighs the entropy of mixing (violating our assumptions for a regular solution), and the free energy curve for these ordered phases is consequently very narrow. The resulting compounds have a negligible range of non-stoichiometry, and are often referred to as *line compounds* due to their negligible width on a phase diagram. Figure 11 illustrates a hypothetical binary phase diagram containing the compounds *AB* and $AB_2$, which are congruently and incongruently melting respectively.

Finally, recalling the case mentioned earlier of an endothermic enthalpy, we note that solid solutions may not be thermodynamically stable for all temperatures. Figure 12 shows an example of *unmixing* driven by a positive enthalpy of mixing of the solid phase. Such cases are essentially driven by stronger chemical bonding between like atoms relative to dissimilar atoms ($E_{AB} < (E_{AA} + E_{BB})/2$). Similar effects can also occur for liquid phases, leading to liquid immiscibility.

### 2.2 *The Gibbs phase rule*

The description presented so far accounts for phase separation in mixtures of two or more components. For the case of the binary system, inspection of Figures 9-13 reveals that there are only a limited number of distinct phases in equilibrium for any given composition or temperature. Is there a fundamental principle limiting the number of phases that can be in equilibrium? The answer is yes, and this is embodied in the Gibbs phase rule.

Consider a mixture of *C* components comprising *P* phases in equilibrium. The Gibbs phase rule reveals the number of available degrees of freedom (*F*) left to the system – that is, the number of intensive parameters that can be independently varied while still maintaining equilibrium:

$F = C - P + 2$      (6).

This expression can be readily obtained by considering the number of intensive parameters that are necessary to fully describe a system in equilibrium. To start, we must specify the amount of each component in each phase, requiring *PC* composition variables. However, since the composition of each phase is specified as a relative mole fraction of each component, one parameter per phase is uniquely determined, reducing this to *P*(*C*-1) variables. In addition, in equilibrium the chemical potential of a component must be the same in each phase, such that if the concentration of component $C_1$ is fixed in phase $P_1$, it must be known for all other *P*-1 phases. This additional constraint reduces the number of parameters to *P*(*C*-1) – *C*(*P*-1). And finally, we must specify the temperature and pressure, which are the same for all phases, giving *F* = *P*(*C*-1) – *C*(*P*-1) + 2 = *C* – *P* + 2 (Equation 6). If the pressure is fixed (often the case experimentally) we arrive at the condensed phase rule

$F = C - P + 1$      (7).

The Gibbs phase rule is perhaps best appreciated with a specific example. Figure 13 shows the binary phase diagram of the two component system Pb-Te at 1 Atmosphere [12]. The phase diagram comprises one line compound, PbTe, and a eutectic mixture at approximately 90% Te. We consider three different regions of the phase diagram. The homogeneous liquid (*L*) comprises a single phase (*P*=1); hence *F* = 2-1+1 =2. Consequently two parameters are required to completely describe the system; these are the temperature and composition. In other words, in the liquid phase the system has two degrees of freedom. The region below the liquidus comprises a mixture of liquid (*L*) and solid (PbTe) phases in equilibrium (i.e. *P* =2), and consequently *F* = 2-2+1 = 1.  In this case, only one parameter is required to uniquely describe the system. For example, if we specify the composition of the liquid, then the temperature is uniquely described. (Note that the Gibbs phase rule refers to intensive variables, and not to absolute amounts. Hence, for a given temperature the relative composition of liquid and solid phases is fixed, although the total amount of each phase can be varied keeping the proportion fixed.) Finally, at the eutectic point three phases are in equilibrium (the liquid and the two solid phases PbTe and Te). In this case *F*=2-3+1=0, meaning that the eutectic is uniquely defined at this pressure (i.e. it is represented by a single invariant point on the phase diagram). This last point is important to appreciate; for a binary mixture the maximum number of phases that can be in equilibrium is three. The Gibbs phase rule implies that four phases cannot be in equilibrium in a binary mixture.

### 2.3 *Ternary mixtures*

We have introduced the fundamental principles governing phase equilibria in the context of a two component system, but these concepts can of course be extended to more complex multi-component systems. Using the Gibbs phase rule as a guide, one can readily imagine the allowed equilibria in ternary mixtures and beyond, although graphically representing these becomes increasingly more challenging. We briefly review the case of a ternary mixture below.

For a ternary mixture, the relative mole fractions of components *A*, *B* and *C* are given by $X_A$, $X_B$ and $X_C$, where $X_A + X_B + X_C = 1$. Just as the composition of a binary mixture can be represented by a point along a line, the composition of a ternary mixture can be represented by a point on a plane, since specification of the relative amount of any two of the components uniquely defines the third. Ternary compositions are typically represented by points on an equilateral triangle (sometimes referred to as a *Gibbs triangle*) rather than Cartesian coordinates. The third axis is uniquely defined by the other two, but it is often convenient to label all three compositions. Figure 14 illustrates use of such a Gibbs triangle to represent the composition of several hypothetical ternary compounds in the *A-B-C* system.

Viewing such a Gibbs triangle in perspective, a further vertical axis can be used to plot the free energy of any given phase, which is then a curved surface (Figure 15(a)). Following exactly the same analysis as for the binary mixture, it is straightforward to show that the chemical potential of components *A*, *B* and *C* in a specific phase are given by $\mu_A = G - X_B(\partial G/\partial X_B)_{X_C} - X_C(\partial G/\partial X_C)_{X_B}$, $\mu_B = \mu_A + (\partial G/\partial X_B)_{X_C}$ and $\mu_C = \mu_A + (\partial G/\partial X_C)_{X_B}$. Extending the graphical construction used in Figure 6 for a binary mixture, we see that the chemical potential of components *A*, *B* and *C* in a ternary mixture defines a *plane* that is tangential to the free energy surface (Figure 15(a)).

Changes in the free energy of the various possible phases as the temperature of the ternary system is lowered lead to a similar progression of single and mixed phase regimes as was found for the binary case, with two principle differences. First, the Gibbs phase rule now allows for up to 4 phases in equilibrium at constant pressure. Hence the ternary eutectic will involve 3 solid phases and liquid in equilibrium. Second, for mixed phase regions, equalization of the chemical potential implies that equilibrium compositions are determined by a common tangential *plane* rather than a common tangential line, since for each component the chemical potential must be the same in each of the phases (Figure 15(b)). Two phase regions occur for compositions that fall between two distinct minima in the free energy. In this case a series of tangential planes can be constructed that are common to both phases (imagine rolling a plane across the surfaces created by two distinct minima). For any given temperature, the resulting equilibrium phases can be represented on a Gibbs triangle. *Tie lines* are then drawn which connect compositions of the two phases that are in equilibrium. For three phase regions there is only one possible plane that is tangential to all three free energy surfaces for any given temperature. Hence the *tie triangle* on the associated isothermal section comprises three phases in equilibrium with uniquely defined compositions at a specific temperature. Figure 16 shows isothermal sections through the resulting phase diagram for a simple ternary mixture for which the pure elements have structures α, β and γ, and the binary mixtures (A-B, B-C and C-A) each comprise simple eutectics and no intermediate compounds. The associated tie lines and tie triangles are described in the figure caption.

It is sometimes useful to represent the liquidus surface on a Gibbs triangle, in this case using contours to mark isotherms. The liquidus surface corresponding to the phase diagram shown in Figure 16 is illustrated in Figure 17. In this specific case, the liquidus comprises three distinct regions, corresponding to liquid in equilibrium with $\alpha$, $\beta$ and $\gamma$ respectively. These regions are separated by eutectic valleys, which converge towards the eutectic point (the minimum melting point). The solidification sequence for an arbitrary composition can be readily appreciated using this diagram. For example, if a melt with composition *X* is cooled, initially phase $\beta$ precipitates, and the composition of the liquid follows the direction marked in Figure 17, away from B. When the composition of the liquid reaches the eutectic valley, a eutectic mixture of $\beta$ and $\gamma$ is precipitated, and the composition of the liquid follows the eutectic valley until finally the entire mixture solidifies at the ternary eutectic.

It is also sometimes useful to represent a portion of a ternary phase diagram by a vertical section through the temperature-composition space. Such a *pseudo-binary* cut reveals the equilibrium phases, but since the tie lines may not be parallel to the direction of the cut, this sort of diagram cannot be used to determine the compositions and relative amounts of the various phases. Examples of this type of pseudo-binary phase diagram appear in Figures 23 and 24. Further discussion of ternary and higher order phase diagrams can be found, for example, in ref [13].

Finally, we briefly comment on the experimental determination of phase diagrams. Phase diagrams are typically determined by thermally cycling mixtures with specific compositions and recording melting and freezing events via differential thermal analysis and calorimetry. This is, however, not the main subject of this article, and the interested reader is referred to ref [11] for more practical details.

## 3. Crystal growth from molten solutions

Having introduced the principles of phase equilibria and the construction of the associated phase diagrams, we are now in a position to appreciate some of the techniques used to grow single crystals from a molten solution.

### 3.1 *Solidification sequence*

We start by considering the solidification sequence as a melt is slowly cooled, using the schematic binary phase diagram shown in Figure 11 for illustration. In the case of a congruently melting compound (*AB* in Figure 11), cooling a stoichiometric melt results in a single phase solid sample. If there are no temperature gradients, and if we neglect nuances of heat conduction associated with the latent heat, the entire volume of the liquid freezes simultaneously. As noted previously, freezing is a first order transition, but for the current discussion we neglect issues associated with nucleation, noting only that for some cases the melt can be considerably undercooled.

In contrast, for an incongruently melting compound ($AB_2$ in Figure 11) cooling a stoichiometric melt (red line marked "path 1" in Figure 11) initially results in phase separation yielding a solid phase (*AB*) in equilibrium with a liquid phase with the composition given by the liquidus line on the phase diagram. The composition of the liquid progressively changes as the system is further cooled, following the liquidus line. Below the peritectic temperature the equilibrium phase for this composition is solid $AB_2$. However, kinetics severely limits the ability of the system to transform to the equilibrium phase. Indeed, in isolation solid *AB* is stable at this temperature, so the driving force for a phase transformation is only felt at the edges of any of the solid phase *AB* where the solid is in contact with the liquid with a different composition. Consequently the liquid composition continues to follow the liquidus, now yielding a mixture of liquid in equilibrium with solid $AB_2$. Once the system is cooled to the eutectic temperature, the remaining liquid freezes yielding a eutectic mixture, in this case of solid $AB_2$ + *B*. The final eutectic freezing happens at a single temperature, so the resulting eutectic mixture typically consists of a densely intergrown mixture of the two phases. Hence, the solidification sequence involves precipitation of solid *AB*, followed by solid $AB_2$, followed by the remaining melt solidifying to give a eutectic mixture of $AB_2$ and *B*. More rapid cooling can circumvent some of the intermediate phases, or even stabilize metastable crystalline or amorphous phases, but this is not directly relevant to our discussion of crystal growth of thermodynamically stable compounds.

Still using Figure 11 for illustration, we can also consider the solidification sequence associated with cooling a melt composition that intersects the liquidus associated with phase $AB_2$ (blue line marked "path 2"). In this case, the first solid phase to precipitate is $AB_2$, followed by eutectic solidification at the eutectic temperature.

### 3.2 *Common techniques*

The above analysis presents clear methodologies for growing single crystals of congruently and incongruently melting compounds from molten solutions.

For congruently melting compounds, one can of course cool a stoichiometric melt. Since the entire system nominally freezes simultaneously, this typically results in a densely intergrown polycrystalline ingot. Depending on the size of the individual crystallites, single crystals can sometimes be mechanically separated, but this is not always easy. Use of a temperature gradient enables directional solidification, and use of a tapered crucible can help control nucleation, such that a large single crystal can be grown under the correct conditions (this is the essence of the *Bridgman* and *Stockbarger* techniques). Equally, a seed crystal can be attached to a cold finger and lowered into a stoichiometric melt close to the melting point. The associated temperature gradient leads to precipitation of additional solid phase. Slowly pulling the seed crystal out of the melt therefore results in a continuous boule, which if the temperature gradients and speed of pulling are accurately controlled can be a single crystal aligned with the initial seed (this is the essence of the *Czochralski* technique). Alternatively, congruently melting compounds can also be grown by cooling a non-stoichiometric melt that intersects the liquidus associated with the compound. If the melt is not cooled below a peritectic temperature, if there is one associated with the specific side of the phase diagram, then a single solid phase results. The remaining liquid can be decanted to separate the crystals. In this case the nucleation is poorly controlled but the resulting crystals grow over an extended range of temperature and are therefore larger and also less strained than crystals that are grown when a stoichiometric melt is cooled without a temperature gradient. Consideration of the lever rule reveals that the rate at which solid precipitates from the melt (with respect to changes in temperature) is inversely proportional to the slope of the liquidus, which can influence the choice of melt composition. In cases for which there is an appreciable width of formation, and for which precise control of the stoichiometry of the solid phase is required, careful consideration of the melt composition and the decanting temperature are required (see the examples of $Bi_2Se_3$ and $Bi_2Te_3$ in Section 4.2), so there is often a balance that dictates the optimal melt composition for a specific system. A slower cooling rate typically results in a smaller number of larger crystals, but the range of available cooling rates is often limited by practical considerations.

For incongruently melting compounds, it is necessary to find a melt composition that intersects the liquidus associated with the desired phase. For some compounds this is simply not possible, but in cases for which there is an exposed liquidus crystals can be grown by slow cooling a melt with the appropriate composition. As mentioned above, the relative amount of the solid phase is determined by the lever rule, and the rate at which material is precipitated depends on the slope of the liquidus (amount per degree cooled) and the cooling rate (degrees per hour). Hence, even for cases where the phase diagram is already established, it is often necessary to experiment in order to find the optimal melt composition and cooling rate to grow crystals of the desired size. In practice, published phase diagrams are often incomplete or slightly inaccurate, and initial experiments are as much concerned with obtaining the correct phase as with optimizing the crystal size and quality.

In cases for which the melting temperatures are excessively high, it can be helpful to use an additional element or compound as a *flux*. Flux growth is essentially an extension of the growth of single crystals from a non-stoichiometric melt, but in this case we deliberately choose an additional component for the melt that is not necessarily incorporated in the resulting crystals, and that lowers the melting point of the mixture. The process of crystal growth still consists of choosing an appropriate composition such that cooling the melt results in entering a two phase region of the appropriate multicomponent phase diagram in which the desired solid phase is in equilibrium with the liquid. The choice of an appropriate flux is governed by a series of practical concerns, delineated in refs [3-10]. We give a few examples in the following section, but note here that considerable experimentation is often necessary to find a suitable flux. Since the process relies on spontaneous nucleation, flux growth often results in large numbers of relatively small crystals, but optimization of the melt composition and the associated temperature profile can yield very respectable samples in many cases. The remaining melt can either be removed by decanting while still liquid, or by chemical etching after solidification if an appropriate etchant can be determined that preferentially etches the solid flux rather than the crystals.

Finally, we comment on the optical *Floating Zone* (FZ) and *Travelling Solvent Floating Zone* (TSFZ) techniques, which enable the growth of large single crystals of congruently *and* incongruently melting compounds respectively from a molten solution in the presence of a strong temperature gradient. A molten zone is maintained by imaging light from halogen bulbs mounted around the growth region. Similar to the Czochralski technique, a seed crystal is introduced to the molten zone and then slowly drawn out, while simultaneously a polycrystalline feed rod is slowly fed into the molten zone at the other side. This technique can be used with a flux (TSFZ), in which case the composition of the molten zone is different from that of the feed rod and the resulting crystal. Because the nucleation is controlled, this can result in very large single crystals. The absence of a crucible can also be advantageous. For more details, see refs [14 & 15].

## 4. Examples

The concepts introduced above are best appreciated by some specific examples. Here we focus particularly on slow cooling, a technique which is widely applicable and is relatively easy to implement. We draw these examples principally from our own experience, and in each case provide a little background to illustrate the thought process that eventually lead to an optimized growth methodology. This is, however, not intended to be an extensive practical guide to the mechanics of crystal growth, nor is it intended to provide an exhaustive list of fluxes that can be used in different situations. For a more practical orientation the interested reader is referred to the references cited for each example, and to the classic books by Wanklyn [3], Elwell and Scheel [4] and Pamplin [5], and the descriptive reviews by Fisk and Remeika [6], Canfield and Fisk [7], Canfield and Fisher [8], Kanatzidis, Pöttgen and Jeitschko [9] and more recently Bulgaris and zur Loye [10].

We start with three examples of binary systems that illustrate some of the factors affecting the growth of incongruently and congruently melting compounds. We then describe four examples of ternary intermetallic compounds – three of which can be grown from a self-flux, and one which requires a true flux growth. We briefly describe the reasoning that lead to the appropriate melt compositions. Finally, we close with examples of two complex oxides that can be grown from a flux, and describe the reasoning that lead to the specific choice of flux in each case.

### 4.1 $R_2Te_{4+n}$ (R = rare earth element)

Binary alloy phase diagrams are available for most, though not all, combinations of elements [12]. This first example illustrates the growth of a family of incongruently melting compounds for which published phase diagrams exist, although they are somewhat inaccurate. The $R_2Te_{4+n}$ family of layered compounds forms for $n$ = 0, 1 and 2, and comprises alternating layers of square-planar coordinated Te with corrugated $R$Te blocks. The electronic states at the Fermi level arise from partially filled $p_x$ and $p_y$ orbitals associated with the Te sheets. The resulting quasi-2D electronic structure is susceptible to charge density wave instabilities, motivating experiments using single crystal samples that probe the magnetic and electronic properties through the CDW phase transitions (see for example [16]).

The equilibrium binary alloy phase diagram for the representative case of Gd-Te is reproduced in Figure 18 from Massalski [12]. As can be readily seen, all three compounds $GdTe_2$, $Gd_2Te_5$ and $GdTe_3$ are incongruently melting, such that crystal growth via slow cooling of a binary solution requires non-stoichiometric melt compositions. In practice, published phase diagrams are often somewhat inaccurate, and considerable experimentation was required in this particular case to establish appropriate compositions and temperature profiles in order to achieve single phase growths of $R$Te$_2$ and $R_2$Te$_5$. The optimal compositions also depend on the rare earth element. Once the appropriate melt compositions were determined, the growth itself is straightforward and yields well-formed single crystals of all three phases. In all cases, binary mixtures of the elements are held in alumina crucibles and sealed in quartz to prevent oxidation. The mixture is heated to a temperature comparatively far above the reported liquidus in order to achieve a homogeneous solution before slowly cooling. The relatively small difference in peritectic temperatures means that crystals of $R$Te$_2$ and $R_2$Te$_5$ grow over a small interval of temperature. In order to grow a single phase in isolation, the growth must be arrested before crossing the lower peritectic ($R$Te$_2$ and $R_2$Te$_5$) or eutectic temperature ($R$Te$_3$) and the remaining melt decanted. In practice this is best accomplished using a centrifuge, following the procedure described by Fisk and Remeika [6]. Further details for the specific compounds can be found in refs [17&18] for $R$Te$_2$, refs [19 & 20] for $R$Te$_3$, and refs [18 & 21] for $R_2$Te$_5$. As a brief commentary, we note that tellurium has a high vapor pressure, boiling at atmospheric pressure at 988 C. The vapor pressure above the binary melt, however, is substantially reduced due to chemical bonding with the rare earth element, and hence solutions comprising a reasonably high rare-earth composition can be safely taken to temperatures exceeding 1100C without rupturing the quartz ampoule. This is particularly necessary for growing crystals of the ditelluride for some members of the rare earth series for which the liquidus temperature is especially high [17,18].

### 4.2 $Bi_2Se_3$ and $Bi_2Te_3$

This second example illustrates the case of congruently melting binary compounds. Both Bi$_2$Se$_3$ and Bi$_2$Te$_3$ are nominally semiconductors, but due to a finite width of formation they readily become degenerate and considerable care must be exercised in order to obtain a composition as close as possible to the correct stoichiometry. Both materials have been of interest for several decades for their thermoelectric properties. More recently they were predicted and found to be topological insulators (see for example reviews by Hasan and Moore [22] and Qi and Zhang [23]), reinvigorating fundamental research in to their electronic properties. Although in theory topological insulators have an insulating bulk and a conducting surface, in practice chemical imperfection leads to finite bulk conductivity for all currently known examples. In order to access the novel physics associated with the surface states of these materials, the bulk carrier density must be carefully controlled.

Both Bi$_2$Se$_3$ and Bi$_2$Te$_3$ are congruently melting (Figure 19 (a) and (b)), and so can be grown by a variety of techniques. Since the materials are nominally semiconductors, subtle deviation from perfect stoichiometry can have a large effect on the electronic properties. Extensive experiments over the last 40 years have sought to determine melt compositions that can yield $n$ or $p$ type materials from binary melts by controlling the nature of the defects. The predominant defects in Bi$_2$Te$_3$, Bi$_2$Se$_3$ and Sb$_2$Te$_3$ are anti-site, A-on-B defects (A$_B$) and vacancies (V$_A$), though interstitial defects also exist. The predominant difficulty with controlling the carrier type is that each defect has a different energy of formation in a given crystal. In Bi$_2$Se$_3$, the defects are primarily V$_{Se}$, which dope $n$-type [24]. This can be mitigated by growing the materials in a Se rich environment, but this will usually increase the formation of Se$_{Bi}$ defects which also dope $n$-type. Similarly, Sb$_2$Te$_3$ is commonly found $p$-type due to V$_{Sb}$ and Sb$_{Te}$ [25], and the same difficulty arises when adjusting the melt composition to control the carrier type. Only in Bi$_2$Te$_3$, where mutual species anti-site and vacancy defects form in approximately equal number [26], can the carrier type be tuned by simply adjusting the melt composition (see Figure 20 and [27]). Slowly cooling the binary melt will also reduce the number of defects and in general lower the carrier density in $Sb_2Te_3$ or $Bi_2Se_3$, but in the case of $Bi_2Te_3$ this can also change the carrier type. However, differences in the nature of the preferred defects in each compound have also proven useful, changing the carrier type by alloying these materials with each other. This process has a combined effect of reducing the number of the original defects, but in addition creating new (carrier compensating) defects. A good example is the alloy $(Bi_{1-x}Sb_x)_2Se_3$. For high carrier concentrations ($10^{19}$ cm$^{-3}$) small amounts of Sb reduce the density to $10^{18}$ cm$^{-3}$ but simultaneously increase the mobility [28]. However, when the unalloyed compounds begin with a much lower carrier concentration of ~$10^{17}$ cm$^{-3}$, small amounts of Sb may reduce the carrier density by a factor of 10, but this also reduces the mobility by a factor of 100 due to the increased defects and decreased Thomas-Fermi screening [29].

### 4.3 $Bi_2Ir_2O_7$

The third example is of an incongruently melting complex oxide that is found in the binary phase diagram of two constituent oxides, in this case $Bi_2O_3$ and $IrO_2$. A binary phase diagram exists, guiding initial crystal growth experiments. Note that for constant valences of the cations, the oxygen content is fixed, and hence the number of independent components in the mixture is still technically two ($BiO_{1.5}$ and $IrO_2$).

$Bi_2Ir_2O_7$ adopts the pyrochlore structure and is isovalent with the well-studied rare earth (*R*) iridates $R_2Ir_2O_7$. These materials have been predicted to harbor various exotic electronic states, including a topological insulator [30] and a Weyl semi-metal [31], motivating considerable recent attention aimed at elucidating their coupled magnetic and electronic properties. Whereas most of the rare earth iridate pyrochlores are Mott insulators, $Bi_2Ir_2O_7$ is in fact metallic [32], enabling study of the electronic properties close to the magnetic instability [33].

Fortunately, the binary phase diagram $Bi_2O_3$-$IrO_2$ has been determined and can be found in the series of oxide phase diagrams published by the American Ceramic Society [34]. We reproduce this phase diagram in Figure 21. Noting that $Bi_2Ir_2O_7$ is incongruently melting, single crystals can be grown by slow cooling a $Bi_2O_3$-rich melt. The melt can be held in a platinum crucible in air, and the remaining liquid decanted at a temperature above the eutectic, revealing well-formed single crystals with an octahedral morphology suitable for detailed investigation of their magnetic and electronic properties [35].

### 4.4 $CeCoIn_5$

We now move our attention to ternary systems, for which there are far fewer phase diagrams available. This particular example illustrates a simple strategy for the growth of a ternary intermetallic compound which contains a relatively large proportion of a low melting point element, in this case indium. $CeCoIn_5$ is of interest as a heavy fermion superconductor, providing the opportunity to explore the magnetic and electronic properties of a material near a magnetic instability and raising questions as to the possible role played by spin fluctuations in the superconducting pairing mechanism [36,37]. The structure is a layered variant derived from cubic CeIn$_3$, with alternating CoIn$_2$ layers, and indeed other members of the homologous series also exist, including CeCo$_2$In$_8$. Attention was first drawn to the family of compounds by Petrovic, Fisk and coworkers, who also described crystal growth via slow cooling of a ternary melt [36,38].

In this case, no ternary alloy phase diagram existed to guide the initial crystal growth experiments. However, since the material is rich in indium and since no other ternary compounds are known that lie between CeCoIn$_5$ and In (see Figure 22), it is natural to try slow cooling a ternary melt with an excess of In (i.e. using indium as a *self-flux*). Following the protocol developed by Petrovic and coworkers, equal mixtures of Ce and Co are combined with an excess of In. The mixture can be held in an alumina crucible, sealed in quartz to prevent oxidation. After slow cooling, the remaining liquid can be decanted using a centrifuge. The optimal melt composition for the growth of large crystals is then determined empirically. Further details can be found in the original papers by Petrovic and coworkers [36,38].

### 4.5 *R*$_9$Mg$_{34}$Zn$_{57}$ and *R*$_{10}$Mg$_{40}$Cd$_{50}$

These two closely related examples illustrate the case of ternary intermetallic compounds which can be grown from a self-flux, but for which the appropriate melt composition does not constitute a simple excess of one of the elements, in contrast to CeCoIn$_5$ described above. In the case of *R*$_9$Mg$_{34}$Zn$_{57}$, part of the ternary phase diagram has been determined, providing a useful starting point for the growth of single grains. For the case of *R*$_{10}$Mg$_{40}$Cd$_{50}$ no such phase diagram exists, and considerable experimentation was necessary to determine the appropriate melt composition and temperature profile. Both compounds are quasicrystals – hence the awkward ratios of elements, expressed here as percentages, which do not reflect decoration of a simple unit cell. The materials are of interest because they allow investigation of the effect of quasi-periodic order on the dynamics and eventual freezing of local magnetic moments courtesy of the rare earth (*R*) ions [39,40].

It is still a matter of debate as to whether quasiperiodic order is the minimum energy configuration at T = 0, but nevertheless these phases do appear to be thermodynamically stable at finite temperatures, and equilibrium phase diagrams can be constructed based on standard thermal analysis. In the case of *R*-Mg-Zn, such an analysis yielded the pseudo-binary cut reproduced in Figure 23 [41]. As can be seen, for a reasonably wide range of compositions the icosahedral phase is in equilibrium with the liquid, presenting a natural avenue for the growth of single grain samples via slow cooling of a ternary melt. The optimal composition for growth of large single grains depends on the rare earth and was found by varying the melt composition and temperature profile using this published phase diagram as a starting point. In this particular case, the high vapor pressure of Mg requires that the melt be sealed in Ta crucibles. Separation of the resulting quasicrystals from the remaining melt can still be achieved by decanting, in this case incorporating a Ta strainer inside the sealed Ta crucible [8]. Details of the growth can be found in [42].

In the case of R-Mg-Cd, no phase diagram was available. However, the relatively small concentration of rare earth and the low melting point of the Mg-Cd binary both suggest that a melt rich in Mg-Cd might intersect the liquidus surface associated with the desired phase, and indeed after some experimentation this was found to be the case. Details of the growth can be found in [43].

**4.6 $BaFe_2As_2$**

$BaFe_2As_2$ is a prototypical antiferromagnetic "parent" compound of the iron pnictide superconductors (see for example [44-46]). This family of compounds is of deep interest not only in terms of understanding the interplay of various broken symmetry states and superconductivity, but also for the perspective that the materials offer on the older and perhaps more subtle problem of understanding the origin of high temperature superconductivity in the cuprates.

Early work from Zimmer and coworkers [47] showed that the closely related quaternary compound LaFePO can be grown from a Sn flux, suggesting that a similar strategy might also work for $BaFe_2As_2$. This turns out to be true, but it was quickly discovered that there is an appreciable solid solubility of Sn in $BaFe_2As_2$ which strongly affects the intrinsic properties [48] and it is preferable to grow single crystals from a ternary melt instead. (The converse is apparently true for the closely related case of $CaFe_2As_2$, due to a temperature-dependent width of formation [49].) The pseudo-binary cut of the Ba-Fe-As ternary phase diagram reproduced in Fig 24 from Morinaga *et al* [50] makes it clear that single crystals of $BaFe_2As_2$ can indeed be grown from an excess of FeAs (i.e. a self-flux), though in practice this phase diagram was determined after the first reports were published. Various methods have been employed, including Bridgman [50] and slow cooling techniques. In the latter case, the resulting single crystals can be separated from the remaining melt by decanting at a temperature above the ternary eutectic. Various groups including our own [51] have described essentially similar protocols for this growth following the initial reports by Sefat *et al* [52], Luo *et al* [53], and Wang *et al* [54].

Although this article is not intended to be a practical guide, nevertheless a note of caution is appropriate here given the extreme toxicity of arsenic. In particular, we note that arsenic sublimes at a relatively low temperature (615 C). Hence a two-step process is favored, in which first FeAs is produced by reacting elemental Fe powder and As. The mixture is slowly heated in a sealed ampoule, producing polycrystalline FeAs through reaction of the Fe with As vapor (see for example refs [53] and [55]). To avoid rupture of the quartz ampoule containing the reagents it is important to avoid heating rapidly through the sublimation temperature. The resulting FeAs is then mixed in the appropriate ratio with Ba for the crystal growth of $BaFe_2As_2$, as described by several authors [51,52,54,55]. To minimize risk of rupture, care must be exercised in sealing the quartz ampoule containing such reactions. In addition, the furnace must be housed in an enclosure connected to an adequate fume hood ventilation system as a safety precaution.

**4.7 $Yb_{14}MnSb_{11}$**

In many cases it is preferable or necessary to use a flux to obtain single crystals of the desired phase. This is particularly true for refractory compounds which might otherwise require excessively high temperatures. Factors affecting the choice of flux include the necessity that the various components are all soluble in the flux, and that they do not form an ordered compound with the flux, at least over some range of compositions and temperature. Practical considerations and lists of common fluxes can be found, for example, in [3,4 & 10] for oxides, and in [6-9] for intermetallics. There are many examples that one could cite; here we choose $Yb_{14}MnSb_{11}$ because it provides a natural extension of the discussion of the growth of CeSb from a Sn flux cited in [7] and illustrates the associative method that often guides the highly empirical process of crystal growth. The material is a member of a broader family of Zintl compounds with the same structure, of interest for their magnetic properties and more recently as thermoelectric materials.

Faced with the desire to grow such a material, one has to start somewhere. Although Sb is in many cases a good self-flux (for example in materials such as the rare earth diantimonides [7], or the closely related ternary compounds such as $RAgSb_2$ [56]), nevertheless the relatively high melting point of the binary Yb-Sb compounds argues that these compounds and the associated liquidus might dominate the phase diagram. This is where a certain amount of experience is helpful. It turns out that rare earth monoantimonides can be grown from a Sn flux, as described for example in [7], which is suggestive that Sn might also be a suitable flux for $Yb_{14}MnSb_{11}$. This is indeed the case, and after some experimentation with melt compositions and temperature profiles a very satisfactory procedure can be determined to grow relatively large, well-formed single crystals (see [57] for details). Of course there was no way to know that this would work given the absence of appropriate quaternary phase diagrams, and the reasoning that lead to this choice of flux is hardly rigorous. There are quite possibly other fluxes that might work just as well for this material, and of course for other materials different fluxes will be appropriate. Experience eventually builds up an intuition, or at least contributes to an associative process in which experimental growths for new materials are guided by previous experiences. The following two examples reinforce this perspective in the context of complex transition metal oxides.

## 4.8 $BaCuSi_2O_6$

Similar to the case of intermetallic compounds discussed above, determining a suitable flux for the growth of single crystals of a complex transition metal oxide is far from an exact science. Nevertheless, the enterprising physicist can draw on association and as many other clues as present themselves to guide the empirical process. $BaCuSi_2O_6$ provides an interesting example. Originally used as a manmade pigment "Han purple" [58], the material is also of interest as a quantum magnet for which the field-induced magnetically ordered state shows remarkable analogies to a Bose Einstein condensate [59].

Intriguingly, the original archeological samples of $BaCuSi_2O_6$ contained traces of lead, leading to speculation that the ancient chemists had used lead salts in order catalyze decomposition of the high melting point precursor barite ($BaSO_4$) and lower the melting point of the powder mixtures they used to create the pigment [58]. This evidence was enough to suggest trying PbO as a flux for the growth of single crystals. Remarkably, this works, although it turns out that $LiBO_2$ yields larger crystals with fewer flux inclusions. The growth can be performed in air using polycrystalline $BaCuSi_2O_6$ as a precursor, and the residual flux separated by decanting in a centrifuge; details can be found in refs [60 & 61]. The process that lead to using $LiBO_2$ as a flux was highly empirical and was based largely on previous success at growing large single crystals of the copper orthoborate $Sr_2Cu(BO_3)_2$ from the same flux [62] following earlier work by Smith and Keszler [63].

### 4.9 $Ba_3Mn_2O_8$

As a final example we consider the case of $Ba_3Mn_2O_8$. Similar to $BaCuSi_2O_6$, the material is of interest due to its field induced magnetically ordered states, in this case associated with both triplet and quintuplet states of $MnO_4$ dimers [64,65]. Charge counting reveals a rather unusual formal valence of 5+ for the Mn ions. As a consequence, we sought an oxidizing flux in order to avoid the more common 2+, 3+ and 4+ valences found in the Ba-Mn-O system such as $Ba_2MnO_3$, $Mn_2O_3$ and $BaMnO_3$ respectively. Based on previous experience with transition metal ions in very high formal valences (especially the case of $Ba_2NaOsO_6$ described in [66]), NaOH was tried as a possible flux (NaOH is an oxidizing flux because the molten flux loses water to the atmosphere, leaving behind $Na_2O$ which is highly oxidizing). After some experimentation this was indeed found to work very well using polycrystalline $Ba_3Mn_2O_8$ as a precursor; further details can be found [64 & 67]. The growth is however somewhat different from all of the previous examples in that it proceeds in part by slow cooling, but also in part by evaporation of NaOH. Hence, instead of following a vertical path on a temperature-composition phase diagram, the composition actually changes with time as the flux evaporates. It is possible to control the rate at which the flux evaporates by partially sealing the alumina crucible holding the melt with an alumina cap, which can affect the size of the eventual crystals grown. Residual flux can be removed by soaking in water.

### 5. Concluding remarks

We have presented an introduction to the physical principles underlying crystal growth of thermodynamically stable phases from a molten solution, emphasizing the basics of phase equilibria and focusing on the specific case of slow cooling. Further details of the thermodynamics and kinetics of phase equilibria and phase transformations can be found in standard text books on the subject, including that by Porter and Easterling [1]. Several examples were presented by way of illustration, and the interested reader is directed towards references cited in the main text and the book sections by Wanklyn [3] and Elwell and Scheel [4] (and references therein) for more detailed discussion of practical considerations. One specific aspect that we have not touched upon is the equilibrium between the vapor and solid phases, which is discussed, for example, in [2]. For some transition metal oxides it is especially important to carefully control the partial pressure of oxygen ($P_{O2}$) above the solution in order to stabilize a specific phase and/or valence. For materials with a sufficiently large vapor pressure of the constituents, crystal growth via *physical vapor transport* (PVT) is a possibility. Indeed, although the phase diagram in Figure 13 indicates that single crystals of PbTe can be grown from a molten solution, nevertheless crystals with much improved electronic properties (homogeneity and mobility) can be grown via PVT due to the relatively large vapor pressure of molecular PbTe above the solid phase [68-70]. Materials which have a low vapor pressure can sometimes also be grown from the vapor phase if a suitable transport agent can be found (referred to as *chemical vapor transport*, CVT). An extensive discussion of these vapor phase techniques can be found in Schäfer's excellent monograph [71].

Finally, we note that not all phases of interest are necessarily thermodynamically stable at ambient pressure. Some materials of interest require synthesis at high pressures, and can be quenched to ambient pressure as a metastable phase. Others exist as kinetic products, and must be removed from the furnace after a certain amount of time at temperature in order to quench the metastable phase. Crystal growth of such metastable phases requires considerable experimentation. A good example of the latter is $A$Cu$_4$S$_3$ ($A$ = K,Rb). Ter Haar *et al* [72] have shown that over a fairly wide range of temperatures KCu$_4$S$_3$ and K$_3$Cu$_8$S$_6$ appear for a short period of time as a kinetic product before the thermodynamically stable phase KCu$_3$S$_2$ is obtained. Consequently, relatively large well-formed single crystals of the closely related material RbCu$_4$S$_3$ can be grown by pulling a melt with the appropriate composition from a tube furnace at a specific rate [35]. Too fast of a removal results in microcrystalline samples. Too slow of a removal results in the wrong phase altogether.

**Acknowledgments**

The majority of the examples drawn upon in this article are based on work conducted by past and present students and post-docs at Stanford University, including J. G. Analytis, J. H. Chu, A.S. Erickson, P. Giraldo-Gallo, H. H. Kuo, H. O. Lee, Y. Matsushita, S. C. Riggs, N. Ru, S. E. Sebastian, E. C. Samulon, M. C. Shapiro and K. Y. Shin, for whose spirited dedication IRF is deeply grateful. Two of the examples, R$_9$Mg$_{34}$Zn$_{57}$ and Yb$_{14}$MnSb$_{11}$, date back to formative years spent at the Ames Laboratory, supported by the Department of Energy. IRF particularly thanks P. C. Canfield and coworkers for a delightful ongoing interaction begun at that time. The authors also thank T. H. Geballe, H. H. Kuo, D. Mandrus, S. Risbud, & J. Yan for critical reading of the manuscript and helpful suggestions. Work on R$_2$Te$_{4+n}$, Bi$_2$Se$_3$, Bi$_2$Te$_3$, Bi$_2$Ir$_2$O$_7$ and BaFe$_2$As$_2$ was supported by the Department of Energy, Office of Basic Energy Sciences, under contract DE-AC02-76SF00515. Work on CeCoIn$_5$, R$_{10}$Mg$_{40}$Cd$_{50}$, BaCuSi$_2$O$_6$ and Ba$_3$Mn$_2$O$_8$ was supported by the National Science Foundation, Division of Materials Research, under grants DMR-0134613 and 0705087.

Table 1: Summary of melt compositions and temperature profiles for examples described in this article.

| Material | Flux | Melt Composition | Temperature Profile (°C) | Comments | Reference(s) |
|---|---|---|---|---|---|
| $R\text{Te}_3$ ($R$ = Y, La, Ce, Sm – Tm) | Self-flux | $R_x\text{Te}_{1-x}$ ($x = 0.015 – 0.03$) | $800 – 900 \xrightarrow{96\ hours} 500 – 600$ | Crystal separation by decanting in centrifuge | [19] & [20] |
| $\text{LaTe}_{1.95}$ & $\text{CeTe}_2$ | Self-flux | $R_x\text{Te}_{1-x}$ ($x = 0.14 – 0.18$) | $1150 \xrightarrow{72-120\ hours} 975-1040$ | Crystal separation by decanting in centrifuge | [18] |
| $R_2\text{Te}_5$ ($R$ = Nd, Sm, Gd) | Self-flux | $R_x\text{Te}_{1-x}$ ($x = 0.075 – 0.1$) | $1000 – 1050 \xrightarrow{91\ hours} 880 – 920$ | Crystal separation by decanting in centrifuge | [18] |
| $\text{Bi}_2\text{Te}_3$ | Self-flux | $\text{Bi}_{0.2}\text{Te}_{0.8}$ | $700 \xrightarrow{90\ hours} 475$ | Crystal separation by decanting in centrifuge | [73] |
| $\text{Bi}_2\text{Se}_3$ | Self-flux | $\text{Bi}_{0.4}\text{Se}_{0.6}$ | $750 \xrightarrow{68\ hours} 500$ | Crystal separation by decanting in centrifuge | [29] |
| $\text{Bi}_2\text{Ir}_2\text{O}_7$ | Self-flux | $(\text{Bi}_2\text{O}_3)_x(\text{IrO}_2)_{1-x}$ ($x = 0.33 – 0.4$) | $1100 \xrightarrow{125\ hours} 900$ | Crystal separation by decanting | Unpublished |
| $\text{CeCoIn}_5$ | Self-flux | $\text{Ce}_{0.01}\text{Co}_{0.01}\text{In}_{0.98}$ | $1190 \xrightarrow{200\ hours} 400$ | Crystal separation by decanting in centrifuge | [36] |
| $R_9\text{Mg}_{34}\text{Zn}_{57}$ ($R$ = Y, Tb, Dy, Ho, Er) | Self-flux | $R_x\text{Mg}_y\text{Zn}_{1-x-y}$ ($R \neq$ Tb: $x = 0.03$, $y = 0.51$; $R$ = Tb: $x = 0.026$, $y = 0.574$) | $700 \xrightarrow{100\ hours} 450$ | Crystal separation by decanting in centrifuge | [42] |
| $R_{10}\text{Mg}_{40}\text{Cd}_{50}$ ($R$ = Y, Gd, Tb, Dy) | Self-flux | $R_{0.025}\text{Mg}_{0.275}\text{Cd}_{0.7}$ | $700 \xrightarrow{100\ hours} 400$ | Crystal separation by decanting in centrifuge | [43] |
| $\text{Yb}_{14}\text{MnSb}_{11}$ | Sn | $\text{Yb}_{0.12}\text{Mn}_{0.05}\text{Sb}_{0.09}\text{Sn}_{0.74}$ | $1100 \xrightarrow{130\ hours} 700$ | Crystal separation by decanting in centrifuge | [57] |
| $\text{BaCuSi}_2\text{O}_6$ | $\text{LiBO}_2$ | $(\text{BaCuSi}_2\text{O}_6)_{0.67}(\text{LiBO}_2)_{0.33}$ | $1100 \xrightarrow{225\ hours} 875$ | Crystal separation by decanting | [60] & [61] |
| $\text{Ba}_3\text{Mn}_2\text{O}_8$ | NaOH | $(\text{Ba}_3\text{Mn}_2\text{O}_8)_{0.04}(\text{NaOH})_{0.96}$ | $550 \xrightarrow{60\ hours} 300$ | Crystal separation by etching with water | [64] & [67] |
| $\text{Ba}_2\text{NaOsO}_6$ | NaOH | $(\text{Os})_{0.01}(\text{Ba(OH)}_2)_{0.04}(\text{NaOH})_{0.95}$ | $600 \xrightarrow{132\ hours} 600$ & furnace cool | Crystal separation by etching with water | [66] |

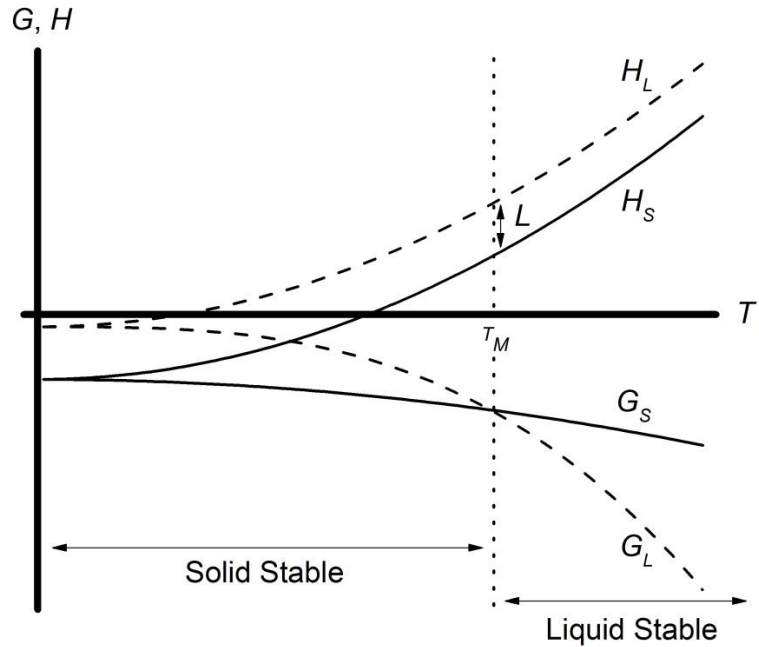

Figure 1: Schematic diagram showing the variation with temperature of the enthalpy (*H*) and free energy (*G*) of the solid and liquid phases of a single component system. The enthalpy rises with increasing temperature due to the increasing internal energy and is always larger for the liquid phase than for the solid phase because the liquid has a larger internal energy. The free energy of the liquid $G_L$ varies more rapidly than that of the solid $G_S$ due to the larger entropy. The melting point $T_m$ is determined by the crossing of the free energy curves, $G_L = G_S$. The latent heat *L* is given by the difference of the enthalpy at $T_m$: $L = H_L - H_S = T_m \Delta S$. The finite latent heat necessarily means that the melting transition is always first order. (Adapted from [1]. Copyright 1992, reproduced by permission of Taylor and Francis Group, LLC, a division of Informa plc.)

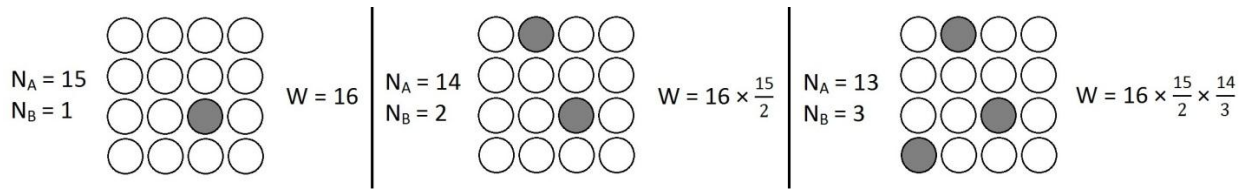

Figure 2: Calculation of the entropy of mixing of two components. For illustration we consider a solid comprising 16 sites. Components A and B are represented by open and filled circles, illustrated for cases $N_B$ = 1, 2 and 3, together with the corresponding number of ways that the atoms can be mixed, *W*. The general case is given by $W = (N_A + N_B)!/(N_A!N_B!)$.

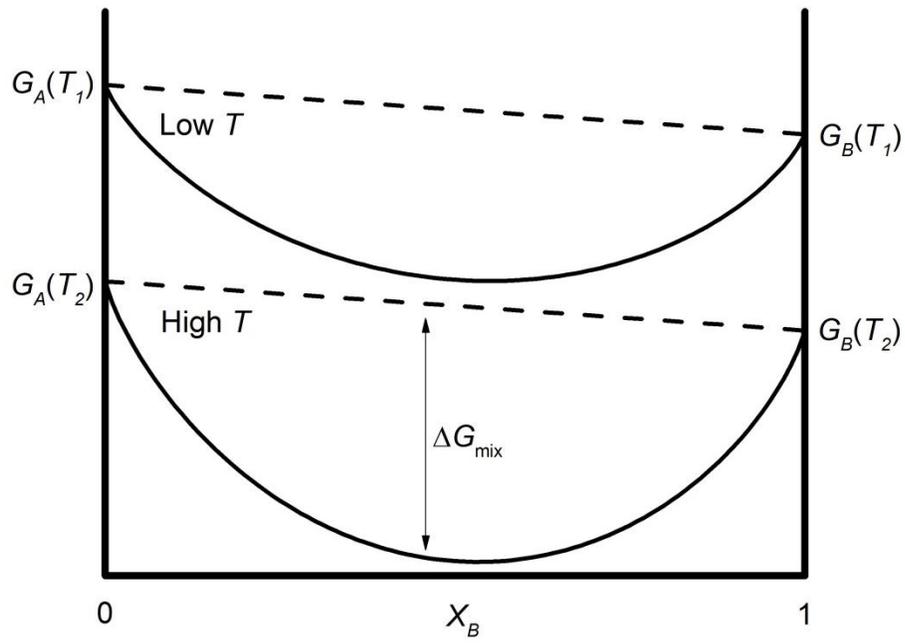

Figure 3: The molar free energy *G* for an ideal solution, showing the change in free energy due to mixing of the two components $\Delta G_{mix}$. The temperature dependence of *G* and $\Delta G_{mix}$ arises from the increased entropy at higher temperatures.

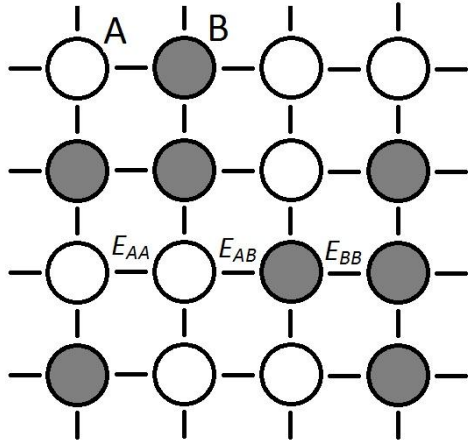

Figure 4: Schematic diagram illustrating bonding in a regular solid solution. The enthalpy of mixing of a regular solution is calculated assuming that the probability of occupancy of lattice sites is random. In this case, the number of chemical bonds between atoms of type $A$, $P_{AA}$, is proportional to the probability that a site is occupied by atom $A$ ($X_A$) multiplied by the probability that an adjacent site is occupied by atom $A$ ($zX_A$), where $z$ is the coordination number. Hence $P_{AA} = zX_A^2 N_{av}/2$ and $P_{BB} = zX_B^2 N_{av}/2$ where the factor of one half avoids double counting. Similarly, the number of bonds between dissimilar atoms is given by $P_{AB} = zX_A X_B N_{av}$. There is no factor of ½ in the expression for $P_{AB}$ because summing over atoms of type $A$ doesn't double count bonds with atoms of type $B$. The energy associated with each type of bond is $E_{AA}$, $E_{BB}$ and $E_{AB}$.

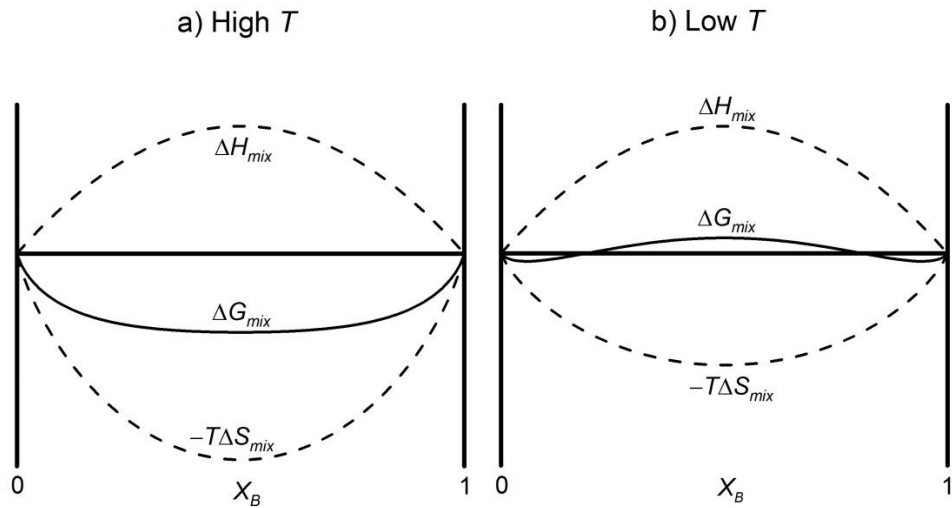

Figure 5: The change in free energy of mixing for an endothermic reaction ($\Delta H_{mix} > 0$). At high temperatures the large entropy leads to a negative $\Delta G_{mix}$ for all compositions. At lower temperatures $\Delta H_{mix}$ is larger than $T\Delta S_{mix}$ for intermediate compositions. However, the larger slope of $T\Delta S_{mix}$ for $X_B \sim 0$ and 1 ensures that at low temperatures, $\Delta G_{mix}$ develops two distinct minima. In this case, intermediate compositions are unstable. The resulting unmixing in solid solutions is referred to as *spinodal decomposition*. (Adapted from [1]. Copyright 1992, reproduced by permission of Taylor and Francis Group, LLC, a division of Informa plc.)

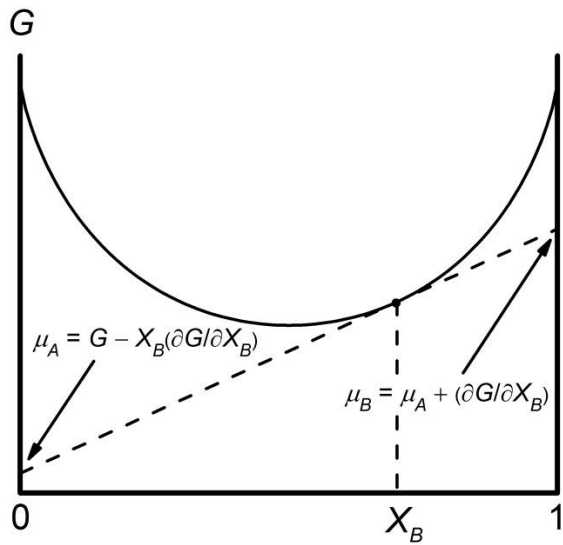

Figure 6: Schematic diagram illustrating the method of intercepts used to determine the chemical potential of components *A* and *B* for a given composition (see main text). Dashed line shows tangent to the free energy curve at the composition $X_B$.

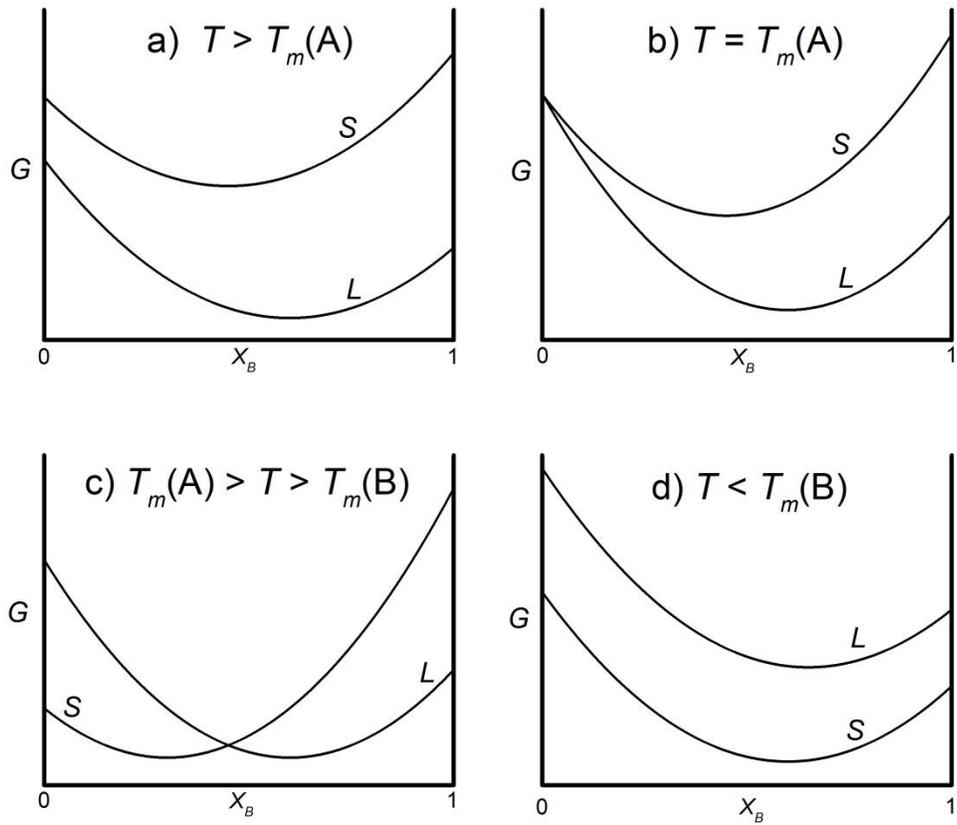

Figure 7: Evolution with decreasing temperature of the compositional dependence of the free energy of solid and liquid phases for an ideal binary solution. (a) $T > T_m(A), T_m(B)$; (b) $T = T_m(A)$; (c) $T_m(A) > T > T_m(B)$; (d) $T < T_m(A), T_m(B)$.

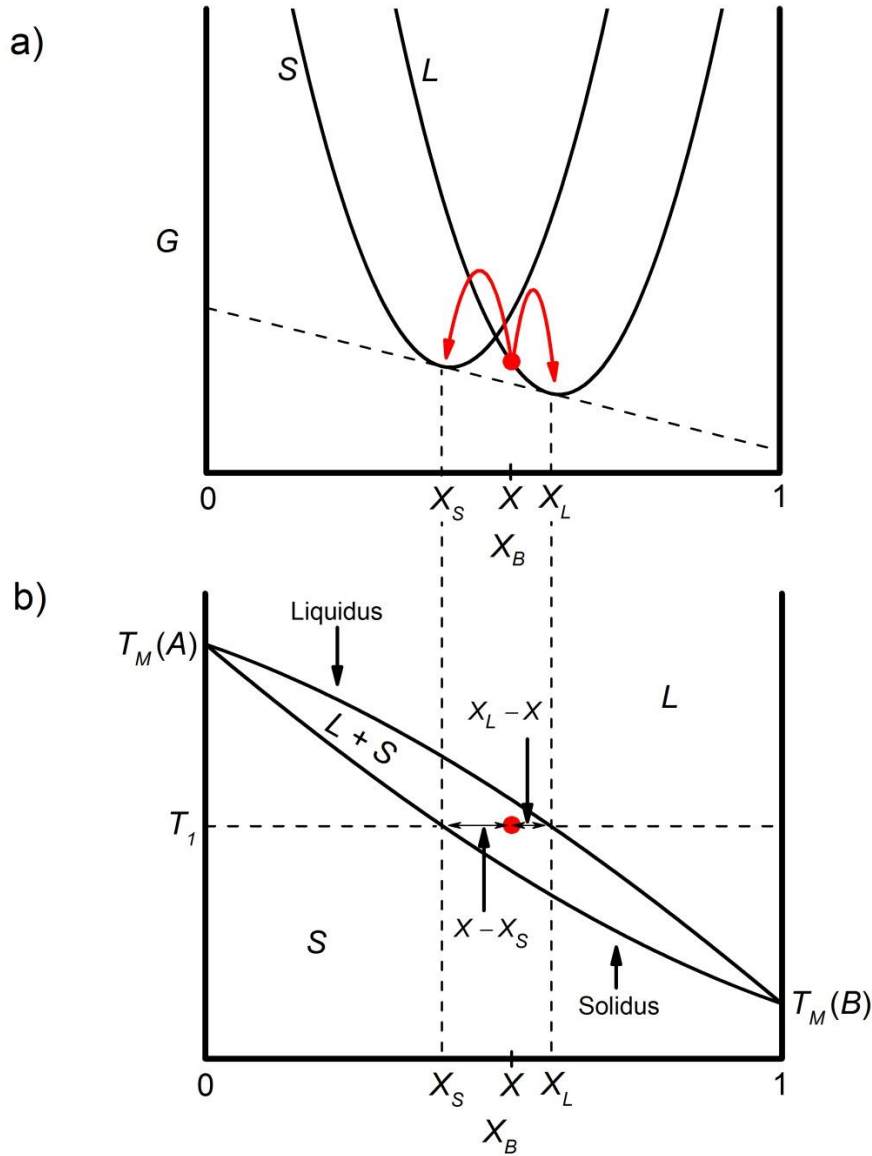

Figure 8: (a) Free energy of solid and liquid phases for an ideal binary solution for a temperature $T_m(A) > T_1 > T_m(B)$. For composition $x$, the total free energy is minimized in equilibrium if the system separates into a mixture of solid ($S$) and liquid ($L$) phases with composition $x_S$ and $x_L$ respectively. The common tangent used to determine these compositions is shown by a dashed line. (b) Construction of the associated phase diagram for the binary mixture of $A$ and $B$. For temperatures between $T_m(A)$ and $T_m(B)$, the *liquidus* and *solidus* lines give the composition of the liquid and solid phases respectively. The relative amount of each phase at any given temperature is given by the lever rule, as described in the main text.

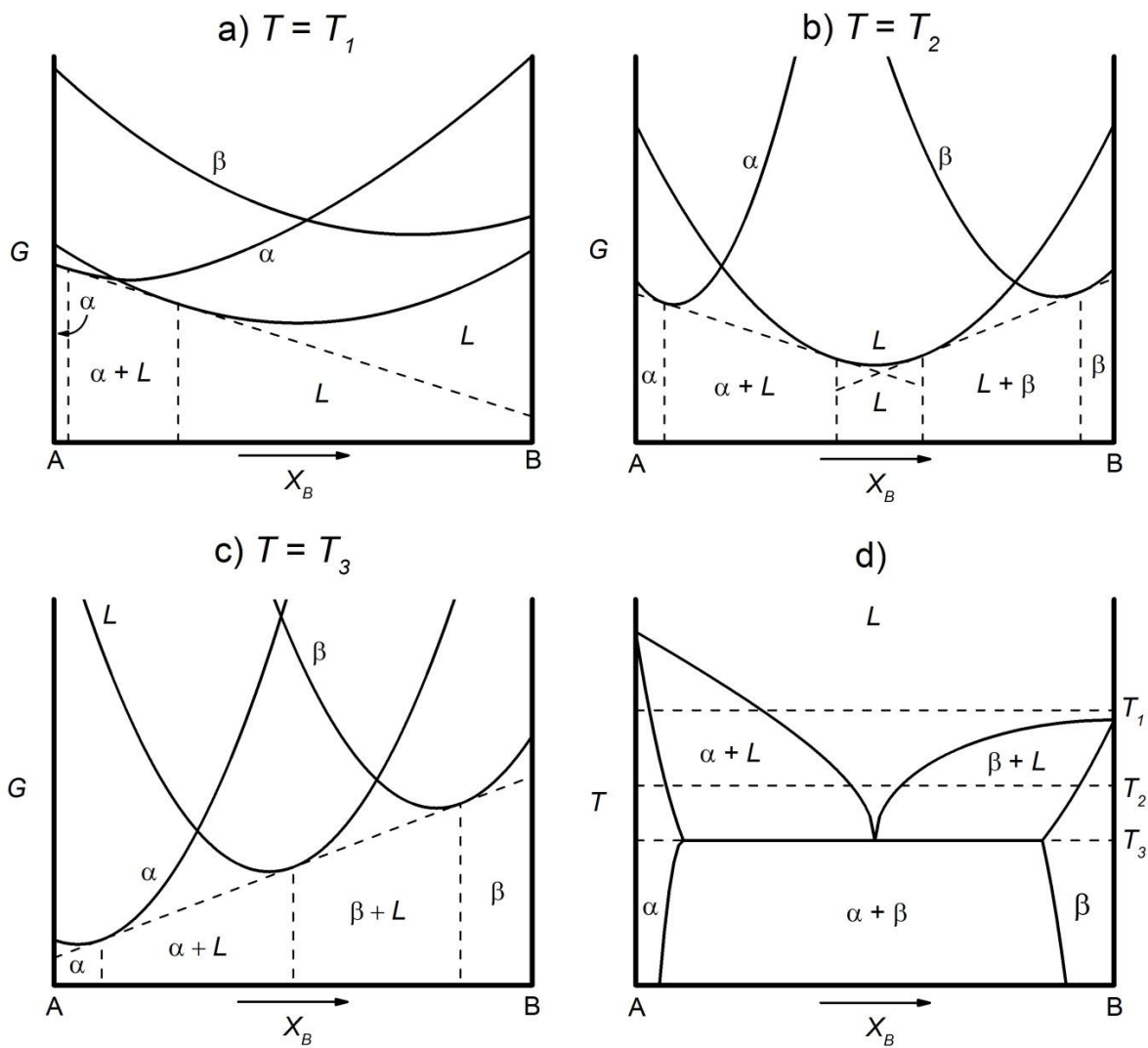

Figure 9: Construction of the phase diagram for a binary mixture for which the end members have different crystal structures. Panels (a) through (c) show free energy curves for the two solid phases, α and β, and the liquid, L, at progressively lower temperatures. Dashed lines show the common tangents used to determine the equilibrium compositions for two-phase regions, and labels indicate the equilibrium phases. Panel (d) shows the corresponding phase diagram. Horizontal lines indicate the temperatures corresponding to panels (a-c). (Adapted from [1]. Copyright 1992, reproduced by permission of Taylor and Francis Group, LLC, a division of Informa plc.)

Figure 10: Schematic diagram showing the construction of the phase diagram for a system with a strong negative deviation from ideality. The large value of $\Delta H_{mix}$ favors an intermediate phase with a different structure ($\beta$) to either of the end members ($\alpha$ and $\gamma$). (a) Free energy of the various phases at a temperature $T_1$. (b) Associated phase diagram. (Adapted from [1]. Copyright 1992, reproduced by permission of Taylor and Francis Group, LLC, a division of Informa plc.)

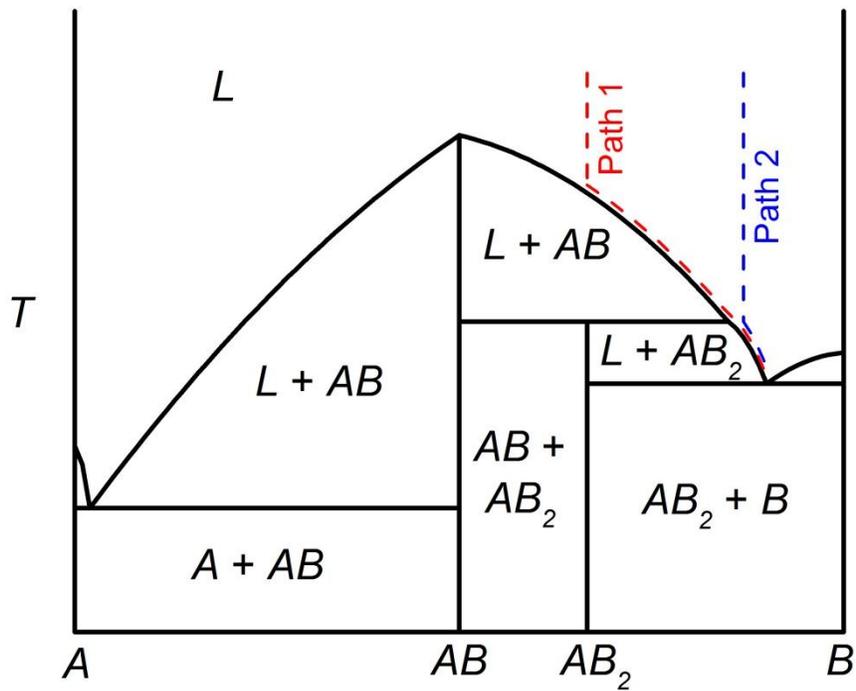

Figure 11: Schematic phase diagram for the hypothetical system *A-B* containing line compounds *AB* and *AB$_2$*. Compound *AB* is *congruently melting*, whereas compound *AB$_2$* melts *incongruently*. Red and blue lines show composition of liquid phase as two different melt compositions are cooled, as described in main text.

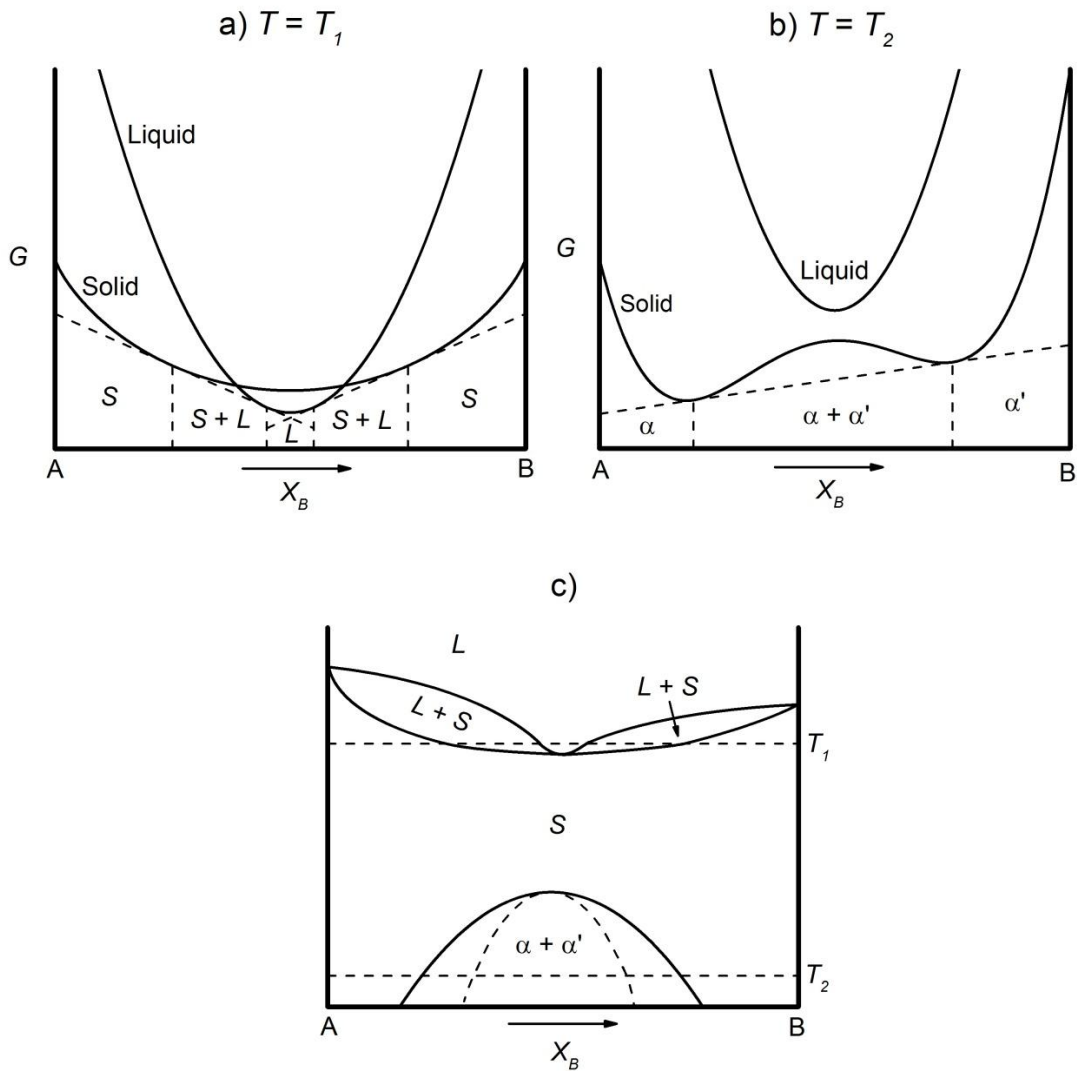

Figure 12: Construction of the phase diagram for a system exhibiting a *miscibility gap*. Panels (a) and (b) show the free energy for specific temperatures in the mixed phase regions, together with the common tangents used to determine the equilibrium compositions. Panel (c) shows the associated phase diagram. The dome-like region delineated with a solid line is known as the *miscibility gap*, and comprises a mixture of solid phases each with the same structure but different compositions, α and α'. The miscibility gap has an inner dome-like *spinodal* region, delineated by a dashed line. Inside the spinodal, the single phase is inherently thermodynamically unstable and there is no barrier to phase separation. However, for compositions between the spinodal curve and the miscibility gap there is a potential barrier for the system to reach the equilibrium state. In this region the single phase is metastable and phase separation proceeds via a process of nucleation. The outer limits of the overall miscibility gap (solid line) are determined by the common tangent construction to the free-energy composition curve at a fixed temperature, while the inflection points (*spinodes*, at which the second derivative of the free energy with respect to composition is zero) of the same curve give the boundaries of the spinodal dome (dashed line). The process of unmixing as a solid solution is cooled is known as *spinodal decomposition*.

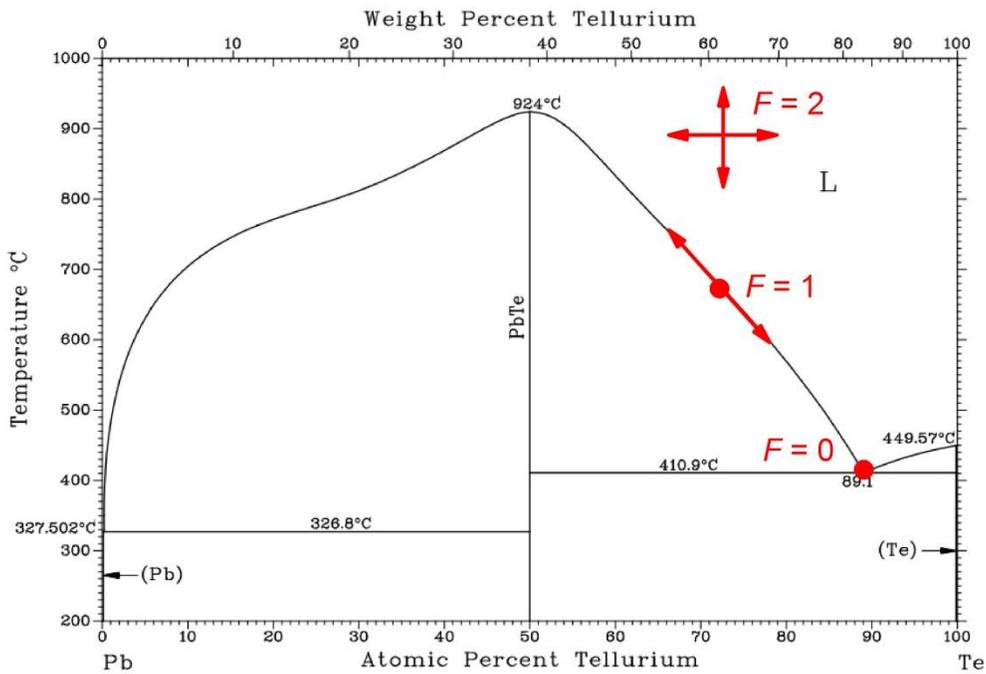

Figure 13: Phase diagram of Pb-Te, after Massalski [12], illustrating the Gibbs phase rule. The number of degrees of freedom ($F$) in the homogeneous liquid phase ($L$) is two, in the mixed phase ($L$ + PbTe) region is one, and at the eutectic is zero. Reprinted with permission of ASM International. All Rights reserved. www.asminternational.org.

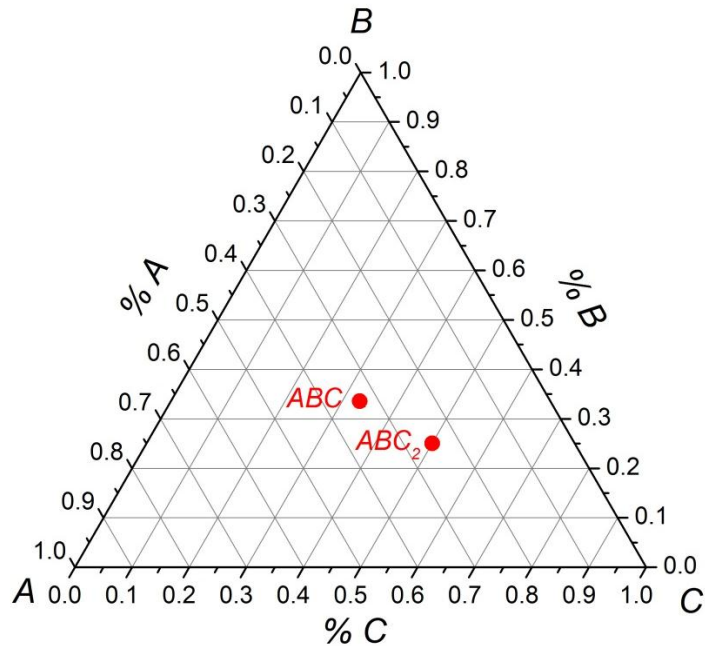

Figure 14: Example of a Gibbs triangle used to represent compositions for a ternary mixture of elements. The position of hypothetical compounds $ABC_2$, corresponding to $X_A$ = 0.25, $X_B$ = 0.25, $X_C$ = 0.5, and $ABC$, corresponding to $X_A$ = 0.33, $X_B$ = 0.33, $X_C$ = 0.33, are marked.

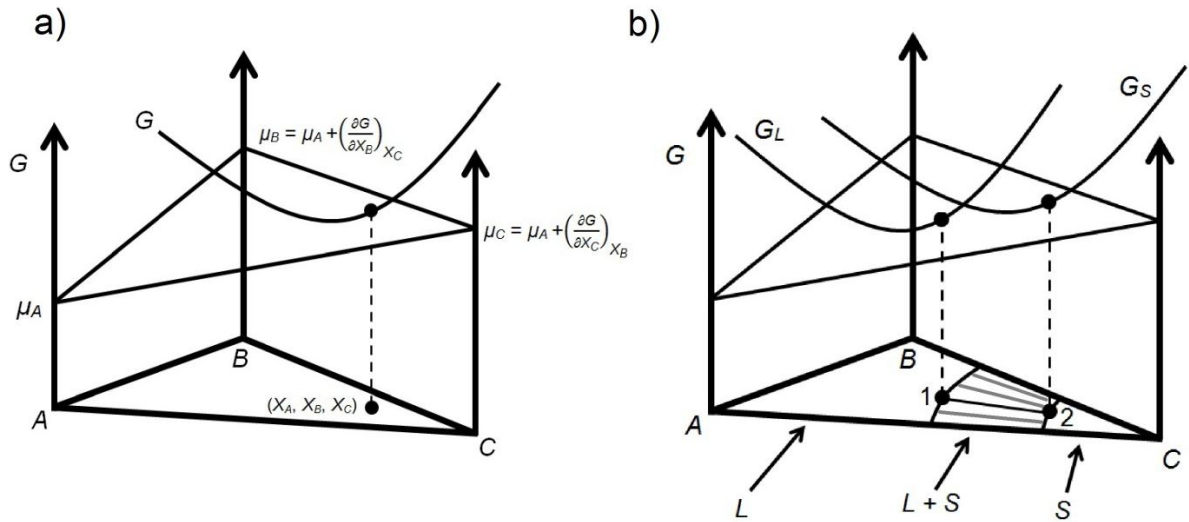

Figure 15: (a) Method of intercepts applied to a ternary mixture. The free energy surface for a particular phase is illustrated. For the composition given by ($X_A$,$X_B$,$X_C$) the free energy has a value $G$. The tangential plane at this composition determines the chemical potential of components $A$, $B$ and $C$, as illustrated. (b) Phase separation occurs for compositions between two minima of the free energy, illustrated here for solid and liquid phases near one corner of the phase diagram. Equalization of the chemical potential for each component in the two phases requires that the equilibrium compositions have a common tangential plane, as illustrated. The equilibrium compositions of the two phases are projected onto the Gibbs triangle, together with the appropriate tie line (dark line between points marked *1* and *2*). This particular plane is one of a set of planes that form common tangents to the two surfaces, leading to a family of tie lines connecting the solid and liquid phases (illustrated in grey).

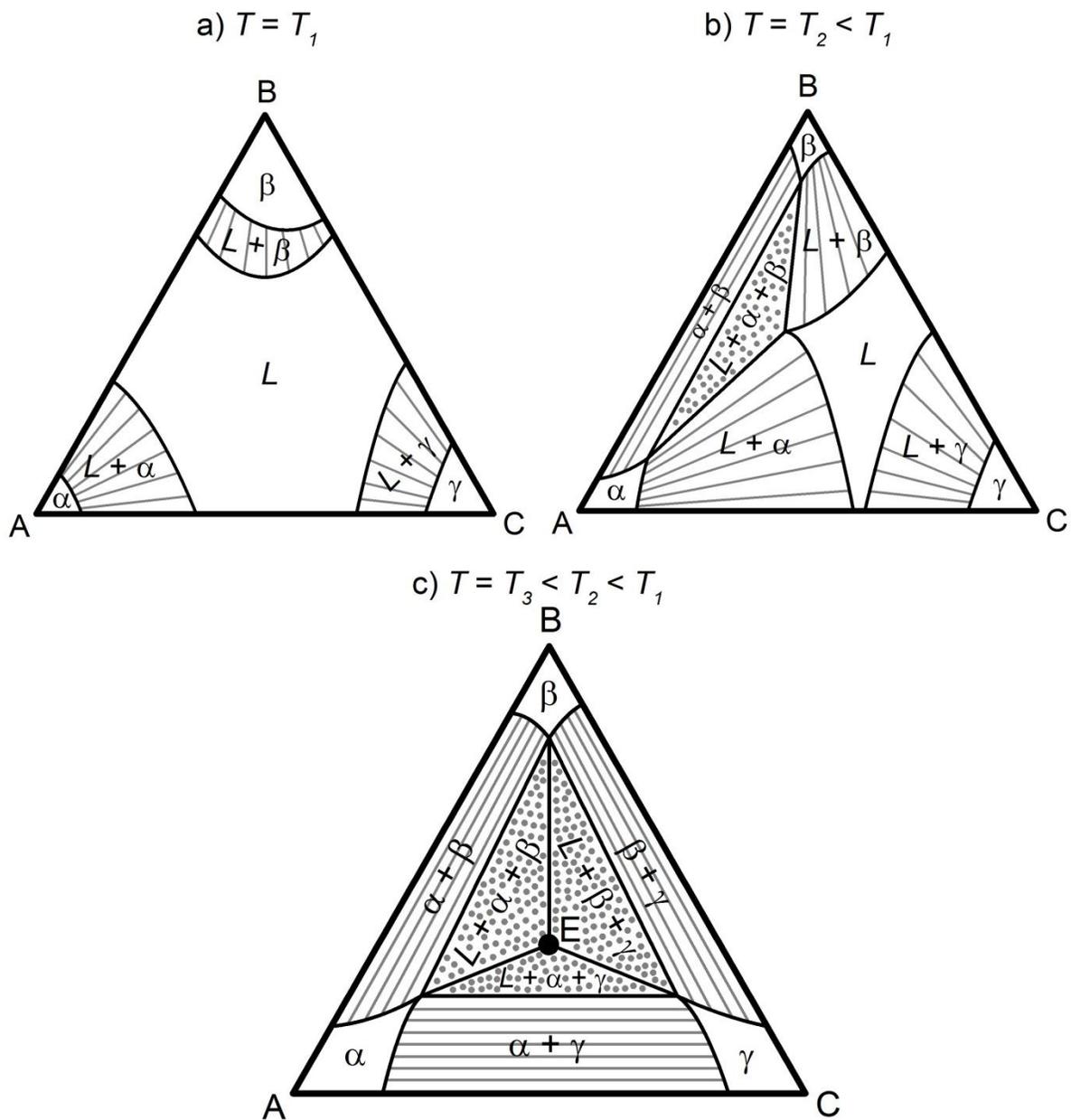

Figure 16: Isothermal sections of the phase diagram for the hypothetical ternary system *A-B-C*, for which the pure elements have structures $\alpha$, $\beta$ and $\gamma$. (a) At a temperature just below the melting point of all three compounds there are three distinct two-phase regions in which the three solid phases are in equilibrium with the liquid (*L*). Tie lines reveal the compositions of conjugate pairs. (b) At a slightly lower temperature the liquid portion of the phase diagram has shrunk in size, and a three phase region is revealed, with liquid in equilibrium with solid $\alpha$ and $\beta$. The composition of the three phases in this tie triangle are given by the corners of the triangle. (c) Further lowering of the temperature eventually leads to a ternary eutectic, marked by the letter *E*, at which three solid phases and the liquid are in equilibrium. The phase diagram at this temperature also comprises three phase regions (dotted tie triangles), two phase regions (with associated tie lines drawn in) and single phase regions (no shading). (Adapted from [1]. Copyright 1992, reproduced by permission of Taylor and Francis Group, LLC, a division of Informa plc.)

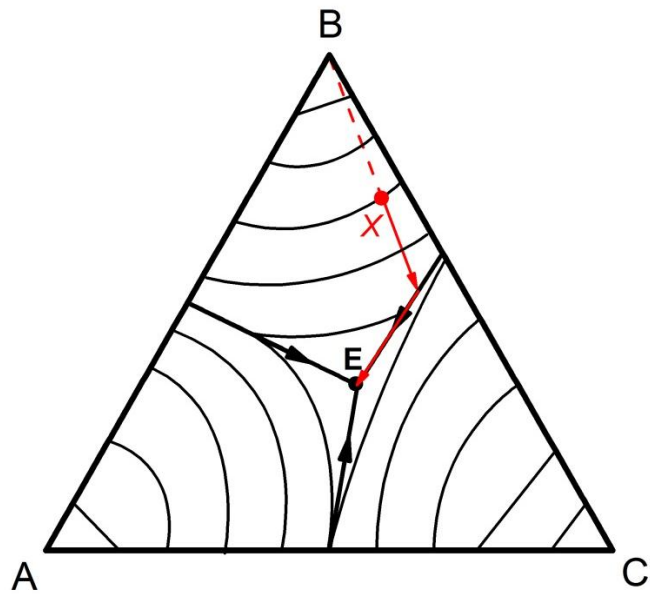

Figure 17: Projection of the liquidus onto the Gibbs triangle for a ternary eutectic system. Contours show isotherms. Solid lines with arrows show eutectic valleys. The ternary eutectic point is labeled *E*. The red line reveals the composition of the liquid as a melt with initial composition *X* is progressively cooled. (Adapted from [1]. Copyright 1992, reproduced by permission of Taylor and Francis Group, LLC, a division of Informa plc.)

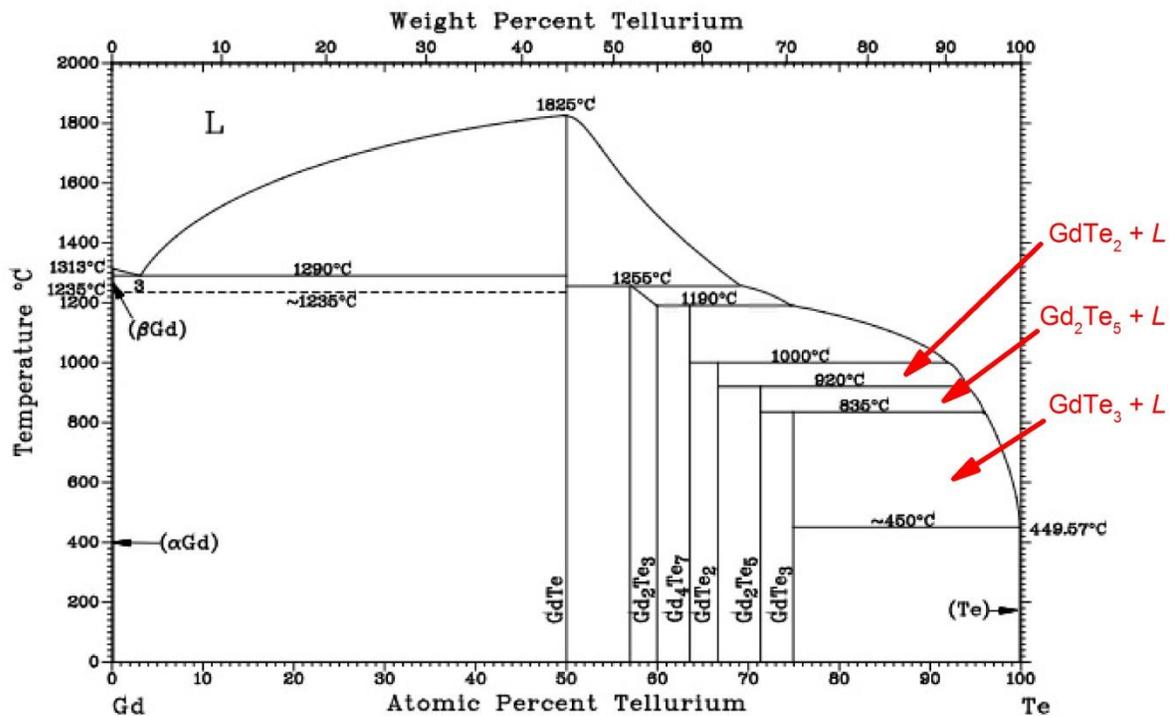

Figure 18: Binary alloy phase diagram for Gd-Te (from Massalski [12].) Two phase regions corresponding to $GdTe_2 + L$, $Gd_2Te_5 + L$ and $GdTe_3 + L$ are labeled, indicating appropriate compositions for crystal growth from a self-flux (excess of Te). In practice, the actual melt compositions required to grow single crystals of $GdTe_2$ and $Gd_2Te_5$ differ slightly from the published phase diagram. Reprinted with permission of ASM International. All Rights reserved. www.asminternational.org.

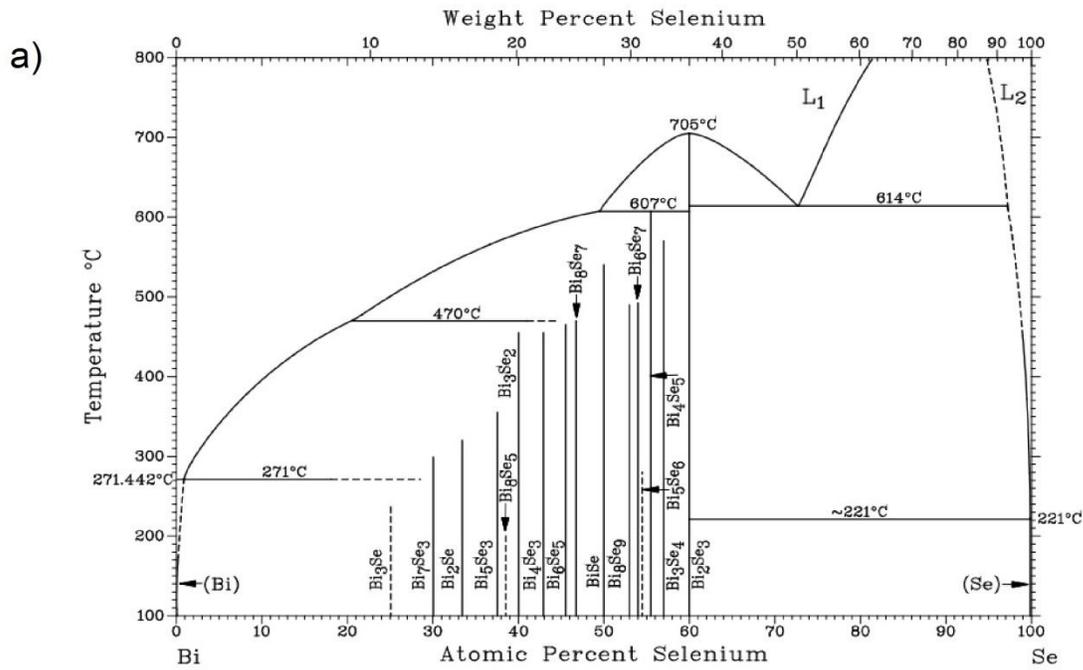
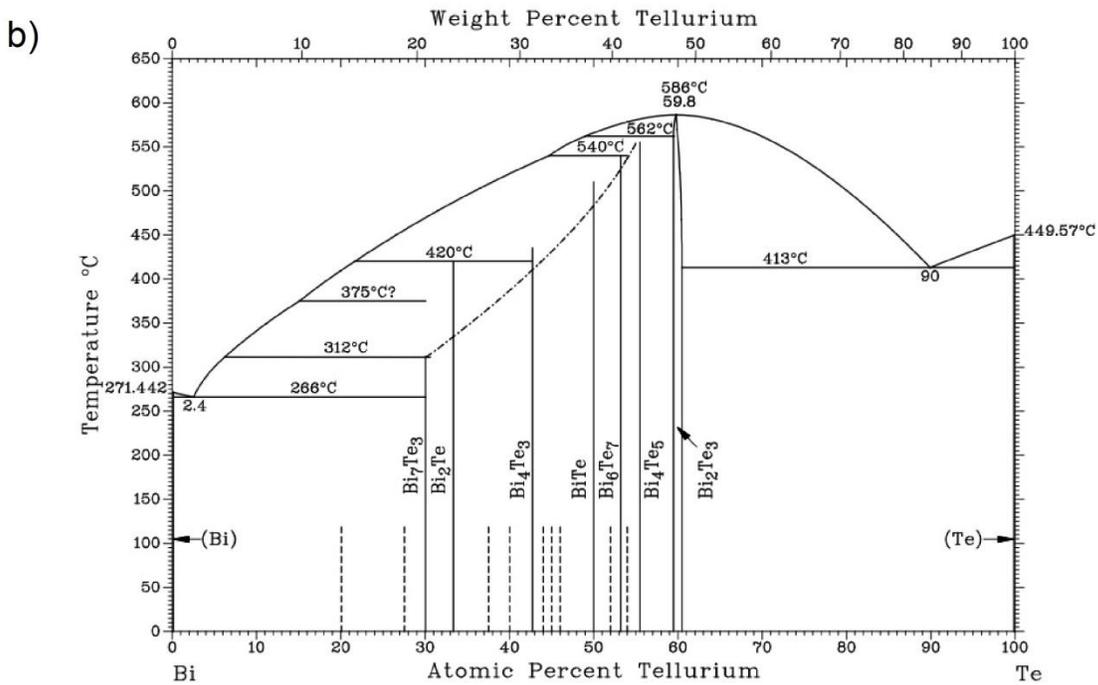

Figure 19: Binary alloy phase diagrams for (a) $Bi_2Se_3$ and (b) $Bi_2Te_3$ (from Massalski [12].) Both compounds are congruently melting. Reprinted with permission of ASM International. All Rights reserved. www.asminternational.org.

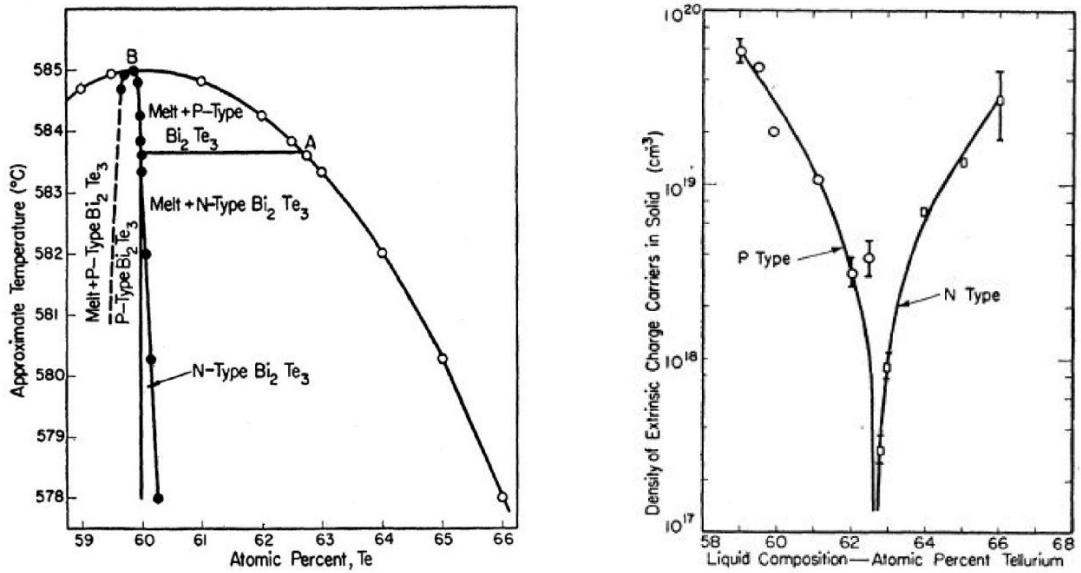

Figure 20: (a) Phase diagram of Bi-Te in the region around $Bi_2Te_3$. (b) Carrier density of the solid phase at 77K, expressed as a function of the original melt composition. Reprinted with permission from [27]. Copyright (1957) by the American Physical Society. http://link.aps.org/abstract/PR/v108/p1164.

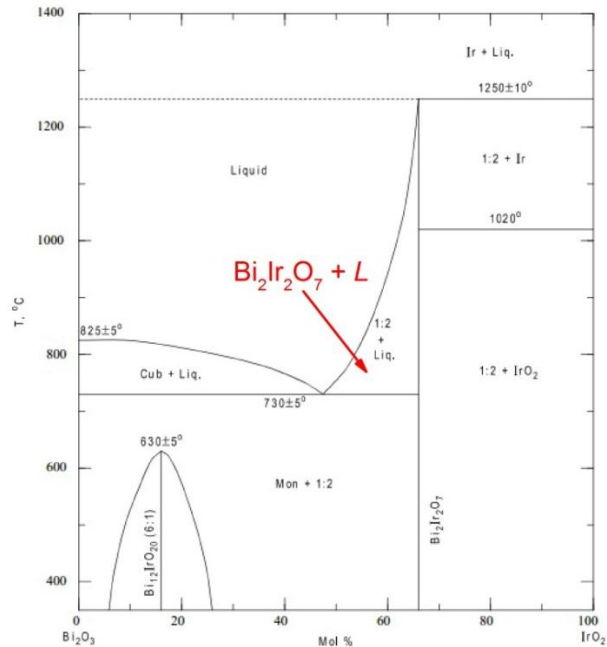

Figure 21: $Bi_2O_3$-$IrO_2$ binary phase diagram, illustrating the two phase region of L + $Bi_2Ir_2O_7$. From Phase Diagrams for Ceramists, Volume XI [34]. Reprinted with the permission of the American Ceramic Society, www.ceramics.org. All rights reserved.

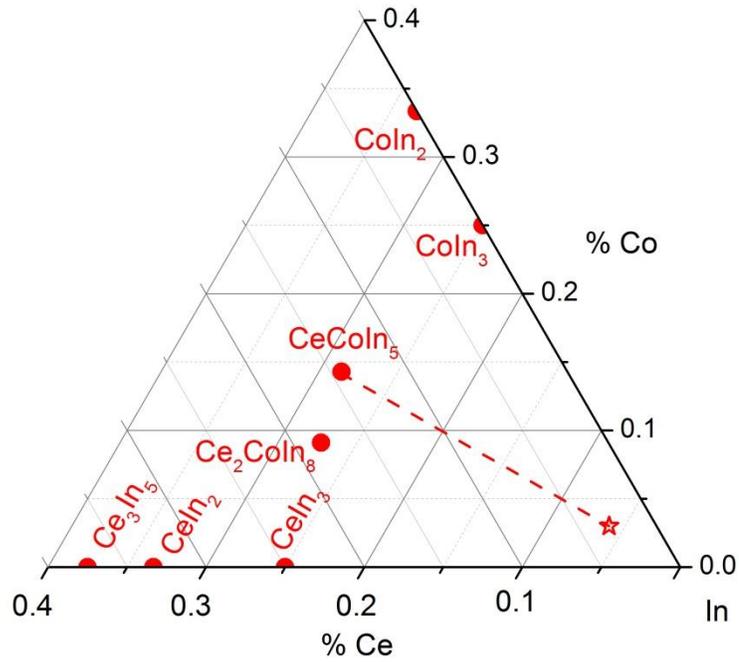

Figure 22: Portion of the Gibbs triangle for the ternary system Ce-Co-In close to In (bottom right corner), illustrating known binary and ternary compounds. The melt composition used to grow single crystals of CeCoIn$_5$ is shown by a star.

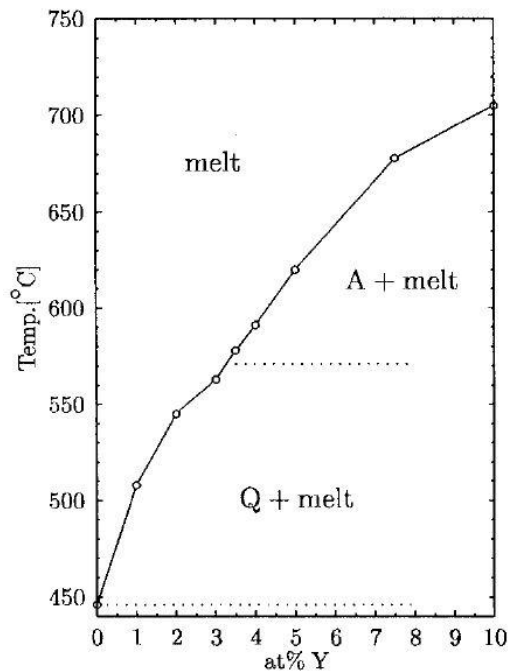

Figure 23: Pseudobinary cut of the Y-Mg-Zn ternary phase diagram for the section defined by (Zn$_{40+2y}$Mg$_{60-3y}$Y$_y$). The two phase region in which the quasicrystalline phase is in equilibrium with the liquid is labeled Q + melt. Reprinted with permission from Langsdorf *et al* [41]. Copyright (1997) by Taylor and Francis.

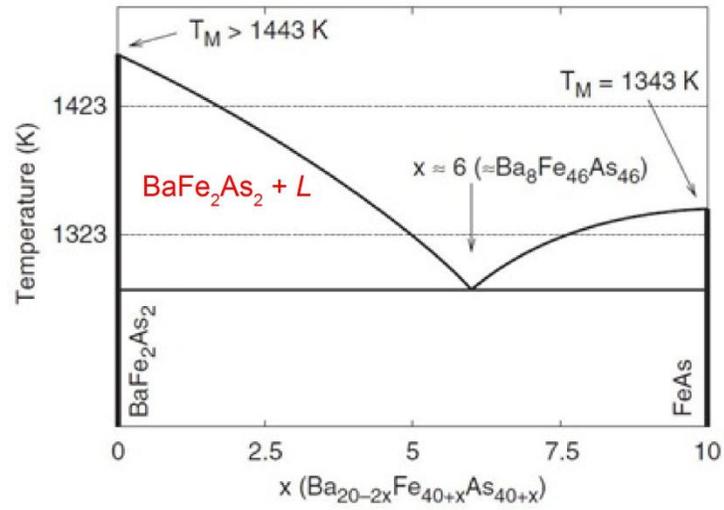

Figure 24: Proposed pseudo-binary phase diagram along the $Ba_{20-2x}$-$Fe_{40+x}As_{40+x}$ line in the Ba-Fe-As system. Reprinted with permission from Morinaga *et al* [50]. Copyright (2009) by the Japan Society of Applied Physics.